\documentclass{eptcs}
\usepackage[utf8]{inputenc}
\usepackage{mdframed,float}
\usepackage{caption}
\usepackage{microtype,amsmath,amsthm,amsfonts,longtable,etoolbox,tikz,colortbl}
\usetikzlibrary{graphs,quotes,decorations.pathreplacing,petri, snakes}
\usepackage{csquotes}
\usepackage{enumitem} 
\usepackage[ruled, linesnumbered, vlined]{algorithm2e}
\usepackage[misc,geometry]{ifsym} 
\usepackage{array}
\preto\tabular{\setcounter{magicrownumbers}{0}}
\newcounter{magicrownumbers}

\newtheorem{example}{Example}
\newtheorem{lemma}{Lemma}
\newtheorem{theorem}{Theorem}

\newcommand{\escale}[1]{\ensuremath{\textbf{\scalebox{0.8}{#1}}}}
\newcommand{\nscale}[1]{\ensuremath{\textbf{\scalebox{0.8}{#1}}}}
\newcommand{\myEdge}[2]{ \tikz[baseline=-1pt]{
\draw[#2,line width=0.3pt] (0,0) -- ++(0.6,0) node[anchor=base, yshift=3pt, pos=0.5] {\escale{$#1$}};
}}
\newcommand{\mylEdge}[2]{ \tikz[baseline=-1pt]{
\draw[#2,line width=0.3pt] (0,0) -- ++(0.9,0) node[anchor=base, yshift=4pt, pos=0.5] {\escale{$#1$}};
}}

\newcommand{\sedge}[1]{ \tikz[baseline=-1pt, photon/.style={decorate,decoration={snake,post length=0.5mm}}]{
\draw[->,line width=0.3pt, photon] (0,0) -- ++(1.2,0) node[anchor=base, yshift=5pt, pos=0.5] {\escale{$#1$}};
}}

\newcommand{\lEdge}[1]{ \tikz[baseline=-1pt]{
\draw[->,line width=0.3pt] (0,0) -- ++(1,0) node[anchor=base, yshift=5pt, pos=0.5] {\escale{$#1$}};
}}

\newcommand{\edge}[1]{\myEdge{#1}{->}}
\newcommand{\ledge}[1]{\mylEdge{#1}{->}}

\newcommand{\fbedge}[1]{\myEdge{#1}{<->}}

\newcommand{\nop}{\ensuremath{\textsf{nop}}}
\newcommand{\inp}{\ensuremath{\textsf{inp}}}
\newcommand{\out}{\ensuremath{\textsf{out}}}
\newcommand{\set}{\ensuremath{\textsf{set}}}
\newcommand{\res}{\ensuremath{\textsf{res}}}
\newcommand{\swap}{\ensuremath{\textsf{swap}}}
\newcommand{\free}{\ensuremath{\textsf{free}}}
\newcommand{\used}{\ensuremath{\textsf{used}}}

\title{On the Parameterized Complexity of Synthesizing Boolean Petri Nets With Restricted Dependency}
\author{Ronny Tredup
\institute{Universit\"at Rostock, Institut f\"ur Informatik, Theoretische Informatik\\ Albert-Einstein-Stra\ss e 22, 18059, Rostock, GErmany \\\email{ronny.tredup@uni-rostock.de}}\\ \and 
Evgeny Erofeev
\institute{Department of Computing Science, Carl von Ossietzky Universit\"at Oldenburg,\\ D-26111 Oldenburg, Germany \\\email{evgeny.erofeev@informatik.uni-oldenburg.de}}
}

\def\titlerunning{Parameterized Complexity of Synthesizing Boolean Petri Nets With Restricted Dependency}
\def\authorrunning{Ronny Tredup and Evgeny Erofeev}
\begin{document}
\maketitle

\begin{abstract}
Modeling of real-world systems with Petri nets allows to benefit from their generic concepts of parallelism, synchronisation and conflict, and  obtain a concise yet expressive system representation. 
Algorithms for synthesis of a net from a sequential specification enable  the well-developed theory of Petri nets to be applied for the system analysis through a net model. 
The problem of $\tau$-synthesis consists in deciding whether a given directed labeled graph $A$ is isomorphic to the reachability graph of a Boolean Petri net $N$ of type $\tau$. 
In case of a positive decision, $N$ should be constructed.
For many Boolean types of nets, the problem is NP-complete. 
This paper deals with a special variant of $\tau$-synthesis that imposes restrictions for the target net $N$: 
we investigate \emph{dependency $d$-restricted $\tau$-synthesis (DR$\tau$S)} where each place of $N$ can influence and be  influenced by at most $d$ transitions.
For a type $\tau$, if $\tau$-synthesis is NP-complete then DR$\tau$S is also NP-complete.
In this paper, we show that DR$\tau$S parameterized by $d$ is in XP.
Furthermore, we prove that it is $W[2]$-hard, for many Boolean types that allow unconditional interactions $\set$ and \textsf{reset}.
\end{abstract}



\section{Introduction}%

Petri nets are widely used for modeling of parallel processes and distributed systems due to their ability to express the relations of causal dependency, conflict and concurrency between system actions. 
In system analysis, one aims to check behavioral properties of such models, and many of these properties are decidable~\cite{DBLP:journals/eatcs/EsparzaN94} for Petri nets and their reachability graphs which represent systems' behaviors. 
The task of system synthesis is opposite: A system model has to be constructed from a given specification of the desired behavior. 
Labeled transition systems serve as a convenient formalism for the behavioral specification, and the goal is then to construct a Petri net whose reachability graph is isomorphic to the input transition system. 
The relevance of the interest to the synthesis is justified in several ways. 
In comparison to the sequential description of the system given by a transition system, the presence of concurrency/parallelism in a Petri net on a fine-grained level allows to encompass the full interleaving in the behavior in a concise yet clear way. 
As a result, this yields a usually much more compact system model without loss of the expressiveness, as long as the synthesis terminates successfully. 
Besides, the alorithms of automatic synthesis ensure that the constructed model is correct-by-design, and hence it does not require any further verification. 
Moreover, the well-developed theory of Petri nets~\cite{DBLP:journals/eatcs/EsparzaN94,DBLP:journals/deds/HollowayKG97} suggests a wide range of methods and techniques for behavioral and structural analysis of the synthesised model, supporting possible improvements and optimization purposes in the initial system.  
Altogether, these allow many areas to benefit from Petri net synthesis,  e.g., extracting concurrency from sequential specifications like TS and languages~\cite{DBLP:journals/fac/BadouelCD02}, process discovery~\cite{DBLP:books/daglib/0027363}, supervisory control~\cite{DBLP:journals/deds/HollowayKG97} or the synthesis of speed independent circuits~\cite{DBLP:journals/tcad/CortadellaKKLY97}.

The complexity of Petri net synthesis significantly depends on the restrictions which are implied by the specification, or imposed on the target system model, or both, and ranges from undecidable~\cite{DBLP:conf/concur/Schlachter16} via NP-complete~\cite{DBLP:conf/concur/TredupR18,DBLP:conf/apn/TredupRW18} down to polynomial~\cite{DBLP:conf/apn/DevillersEH19,DBLP:conf/stacs/Schmitt96}. 

In this work, we study the complexity of synthesis for Boolean nets~\cite[pp.~139-152]{DBLP:series/txtcs/BadouelBD15}, where each place contains at most one token, for any reachable marking.
A place of such a net is often considered as a Boolean condition which is \emph{true} if the place is marked and false otherwise.
In a Boolean Petri net, a place $p$ and a transition $t$ are related by one of the Boolean \emph{interactions}: 
\emph{no operation} (\nop), \emph{input} (\inp), \emph{output} (\out), \emph{unconditionally set to true} (\set), \emph{unconditionally
reset to false} (\res), \emph{inverting} (\swap), \emph{test if true} (\used), and \emph{test if false} (\free).
These interactions define in which way $p$ and $t$ influence each other:
The interaction $\inp$ ($\out$) defines that $p$ must be \emph{true} (\emph{false}) before and \emph{false} (\emph{true}) after $t$'s firing;
$\free$ ($\used$) implies that $t$'s firing proves that $p$ is \emph{false} (\emph{true});
$\nop$ means that $p$ and $t$ do not affect each other at all; 
$\res$ ($\set$) implies that $p$ may initially be both \emph{false} or \emph{true} but after $t$'s firing it is \emph{false} (\emph{true});
$\swap$ means that $t$ inverts $p$'s current Boolean value.

Boolean Petri nets are classified by the sets of interactions between places and transitions that can be applied. 
A set $\tau$ of Boolean interactions is called a \emph{type of net}.
A net $N$ is of type $\tau$ (a \emph{$\tau$-net}) if it applies at most the interactions of $\tau$.
For a type $\tau$, the $\tau$-\emph{synthesis} problem consists in deciding whether a given {\em transition system} $A$ is isomorphic to the reachability graph of some $\tau$-net $N$, and in constructing $N$ if it exists.
The complexity of synthesis strongly depends on the target Boolean type of nets. 
Thus, while $\tau$-synthesis for elementary net systems (the case of $\tau = \{\nop,\inp,\out\}$) is shown to be NP-complete~\cite{DBLP:journals/tcs/BadouelBD97}, the same problem for flip-flop nets ($\tau = \{\nop,\inp,\out,\swap\}$) is polynomial~\cite{DBLP:conf/stacs/Schmitt96}. 

This paper addresses the computational complexity of a special instance of $\tau$-synthesis called \emph{Dependency $d$-Restricted $\tau$-Synthesis} (DR$\tau$S), which sets a limitation for the number of connections of a place.  
This synthesis setting targets to those $\tau$-nets in which every place must be in relation $\nop$ with all but at most $d$ transitions of the net, while the synthesis input is not confined. 
In system modeling~\cite{DBLP:books/daglib/0084790}, places of Petri nets are usually meant as conditions or resources, while transitions are meant as actions or agents. 
Hence, the formulation of $d$-restricted synthesis takes into consideration not only the concurrency perspective but also possible {\it a priori} limitations on the number of agents which compete for the access to some resource in the modeled system. 
From the theoretical perspective, the problem of synthesis has been extensively studied in the literature for the conventional Petri nets and their subclasses, which are often defined via various structural restrictions: 
Recently, improvements of the existing synthesis techniques have been suggested for choice-free (transitions cannot share incoming places)~\cite{DBLP:journals/acta/BestDS18}, weighted marked graphs (each place has at most one input and one output transition)~\cite{DBLP:conf/apn/DevillersEH19,DBLP:journals/fuin/DevillersH19}, fork-attribution (choice-free and at most one input for each transition)~\cite{DBLP:journals/iandc/Wimmel20} and other net classes~\cite{DBLP:conf/ac/Best86a,DBLP:journals/corr/abs-1911-09133}.  
In these works, the limitations were mainly subject to the quantity of connections between places and transitions. 
On the other hand, the results on synthesis of $k$-bounded (never more than $k$ tokens on a place)~\cite{DBLP:conf/apn/Tredup19}, safe (1-bounded) and elementary nets~\cite{DBLP:series/txtcs/BadouelBD15} investigate classes which are defined through behavioral restrictions. 
Further, generalized settings of the synthesis problem for these and some other classes were studied~\cite{DBLP:conf/rp/Tredup19}, and NP-completeness results for many of them were presented. 
In contrast to this multitude of P/T net classes, for Boolean nets, only the constrains for the set of interactions have appeared in the literature, deriving for instance flip-flop nets~\cite{DBLP:conf/stacs/Schmitt96}, trace nets~\cite{DBLP:journals/acta/BadouelD95}, inhibitor nets~\cite{DBLP:conf/apn/Pietkiewicz-Koutny97}. 
This kind of constrain can be considered as behavioral limitation, leaving out the question of synthesis of possible structurally defined subclasses of Boolean nets. 
The present paper aims to piece out the shortage by investigating the notion of $d$-restriction which limits the amount of connections between a place and transitions. 
The notion was initially introduced in~\cite{DBLP:conf/tamc/TE20}, where the complexity of $d$-restricted synthesis has been studied for a number of Boolean types, and the W[1]-hardness of this problem has been proven. 
The current paper extends the previous work and tackles the problem for many types that necessarily include interactions $\res$ and $\set$. 
We demonstrate the W[2]-hardness of $d$-restricted synthesis for these types, which makes a clear distinction to the earlier results. 

The paper is organized as follows. 
After introducing of the necessary definitions in Section~\ref{sct:prelim}, the main contributions on W[2]-hardness of DR$\tau$S are presented in Section~\ref{sct:dnr}.  
Section~\ref{sct:concl} suggests an outlook of the further research directions.
Due to space restrictions, we omit some proofs, which can all be found in the technical report~\cite{DBLP:journals/corr/abs-2007-12372}.

\section{Preliminaries}\label{sct:prelim}%

In this section, we introduce the notions used throughout the paper and support them by~examples.

\textbf{Parameterized Complexity.}	
Due to space restrictions, we only give the basic notions of Parameterized complexity (used in this paper) and refer to~\cite{DBLP:books/sp/CyganFKLMPPS15} for further related definitions. 
A \emph{parameterized} problem is a language $L\subseteq \Sigma^*\times \mathbb{N}$, where $\Sigma$ is a fixed alphabet and $\mathbb{N}$ is the set of natural numbers.
For an input $(x,k)\in \Sigma^*\times \mathbb{N}$, $k$ is called the \emph{parameter}.
We define the size of an instance $(x,k)$, denoted by $\vert (x,k)\vert$, as $\vert x\vert +k$, that is, $k$ is encoded in unary.
Let $f,g:\mathbb{N}\rightarrow \mathbb{N}$ be two computable functions.
The parameterized language $L$ is \emph{slice-wise polynomial} (XP), if there exists an algorithm $\mathcal{A}$ such that, for all $(x,k)\in \Sigma^*\times \mathbb{N}$, algorithm $\mathcal{A}$ decides whether $(x,k)\in L$ in time bounded by $f(k)\cdot\vert (x,k)\vert ^{g(k)}$;
if the runtime of $\mathcal{A}$ is even bounded by $f(k)\cdot\vert (x,k)\vert ^{\mathcal{O}(1)}$, then $L$ is called \emph{fixed-parameter tractable} (FPT).
In order to classify parameterized problems as being FPT or not, the W-hierarchy $\text{FPT}\subseteq W[1]\subseteq W[2]\subseteq \dots \subseteq \text{XP}$ is defined~\cite[p.~435]{DBLP:books/sp/CyganFKLMPPS15}.
It is believed that all the sub-relations in this sequence are strict and that a problem is not FPT if it is $W[i]$-hard for some $i\geq 1$.
Let $L_1, L_2\subseteq \Sigma^*\times \mathbb{N}$ be two parameterized problems.
A \emph{parameterized} reduction from $L_1$ to $L_2$ is an algorithm that given an instance $(x,k)$ of $L_1$, outputs an instance $(x',k')$ of $L_2$ in time $f(k)\cdot \vert x\vert^{\mathcal{O}(1)}$ for some computable function $f$ such that $(x,k)$ is a yes-instance of $L_1$ if and only if $(x',k')$ is a yes-instance of $L_2$ and $k'\leq g(k)$ for some computable function $g$.
If $L_1$ is $W[i]$-hard 
and there is a parameterized reduction from $L_1$ to $L_2$, then $L_2$ is $W[i]$-hard,~too.

\textbf{Transition Systems.}
A (deterministic) \emph{transition system} (TS, for short) $A=(S,E, \delta)$ is a directed labeled graph with the set of nodes $S$ (called {\em states}), the set of labels $E$ (called {\em events}) and partial \emph{transition function} $\delta: S\times E \longrightarrow S$. 
If $\delta(s,e)$ is defined, we say that event $e$ \emph{occurs} at state $s$, denoted by $s\edge{e}$.
An \emph{initialized} TS $A=(S,E,\delta, \iota)$ is a TS with a distinct {\em initial} state $\iota\in S$ where every state $s\in S$ is \emph{reachable} from $\iota$ by a directed labeled path.
\begin{figure}[b!]
\begin{center}
\begin{minipage}{\textwidth}
\begin{center}
\begin{tabular}{c|c|c|c|c|c|c|c|c}
$x$ & $\nop(x)$ & $\inp(x)$ & $\out(x)$ & $\set(x)$ & $\res(x)$ & $\swap(x)$ & $\used(x)$ & $\free(x)$\\ \hline
$0$ & $0$ & & $1$ & $1$ & $0$ & $1$ & & $0$\\
$1$ & $1$ & $0$ & & $1$ & $0$ & $0$ & $1$ & \\
\end{tabular}
\end{center}
\caption{
All interactions $i$ of $I$.
If a cell is empty, then $i$ is undefined on the respective $x$.
}\label{fig:interactions}
\end{minipage}
\begin{minipage}{\textwidth}
\begin{center}
\begin{tikzpicture}[new set = import nodes]
\begin{scope}
\node (0) at (0,0) {\nscale{$0$}};
\node (1) at (1.2,0) {\nscale{$1$}};

\path (0) edge [->, out=-120,in=120,looseness=5] node[left] {\escale{$\nop$}} (0);
\path (1) edge [<-, out=60,in=-60,looseness=5] node[right] {\escale{$\nop$}  } (1);

\path (0) edge [<-, bend right= 30] node[below] {\escale{$\inp, \swap$}} (1);
\path (0) edge [->, bend  left = 30] node[above] {\escale{$\swap$}} (1);
\node () at (0.5,-1.2) {\nscale{$\tau$}};

\end{scope}

\tikzstyle{place}=[circle,thick,draw=blue!75,fill=blue!20,minimum size=6mm]
\tikzstyle{red place}=[place,draw=red!75,fill=red!20]
\tikzstyle{transition}=[rectangle,thick,draw=black!75,
  			  fill=black!20,minimum size=4mm]
			   \tikzstyle{every label}=[red]
\begin{scope}[xshift=6.25cm, node distance=0.5cm,bend angle=45,auto]
\node (x) at (0,0) {};
\node [place] (R1)[tokens =1,left of=x, label=left: \nscale{$R_1$}, node distance =1.5cm]{};
\node [transition] (a)[above of =x]{$a$}
     	 edge [above,] node {$\nscale{\inp}$}  (R1);
\node [transition] (b)[below of =x]{$b$}
     	 edge [ below] node {$\nscale{\nop}$}  (R1);
	 
\node [place] (R2)[right of=x, label=right: \nscale{$R_2$}, node distance =1.5cm]{}
	edge [above] node {$\nscale{\swap}$}  (a)
	edge [below] node {$\nscale{\inp}$}  (b);
	\node () at (0,-1.2) {$N$};
\end{scope}
\begin{scope}[xshift=10.5cm,nodes={set=import nodes}]
		\node (A) at (1.3,-1.2) {$A_N$};
		\node (10) at (0,0) {\nscale{$(1,0)$}};
		\node (01) at (1.3,0) {\nscale{$(0,1)$}};
		\node (00) at (2.6,0) {\nscale{$(0,0)$}};
\graph {(import nodes);
			10 ->["\escale{$a$}"]01;
			01 ->["\escale{$b$}"]00;
			
		};
\end{scope}
\end{tikzpicture}
\end{center}
\caption{%
The type $\tau=\{\nop,\inp,\swap\}$ and a $\tau$-net $N$ and its reachability graph $A_N$.
}\label{fig:type_net_rg}
\end{minipage}
\end{center}
\end{figure}

\textbf{Boolean Types of Nets~\cite{DBLP:series/txtcs/BadouelBD15}.}
The following notion of Boolean types of nets allows to capture {\em all} Boolean Petri nets in a {\em uniform} way.
A \emph{Boolean type of net} $\tau=(\{0,1\},E_\tau,\delta_\tau)$ is a TS such that $E_\tau$ is a subset of the {\em Boolean interactions}:
$E_\tau \subseteq I = \{\nop, \inp, \out, \set, \res, \swap, \used, \free\}$. 
Each interaction $i \in I$ is a binary partial function $i: \{0,1\} \rightarrow \{0,1\}$ as defined in Figure~\ref{fig:interactions}.
For all $x\in \{0,1\}$ and all $i\in E_\tau$, the transition function of $\tau$ is defined by $\delta_\tau(x,i)=i(x)$.
Since a type $\tau$ is completely determined by $E_\tau$, 
we often identify $\tau$ with $E_\tau$.

\textbf{$\tau$-Nets.}
Let $\tau\subseteq I$.
A Boolean Petri net $N = (P, T, f, M_0)$ of type $\tau$ (a {\em $\tau$-net}) is given by finite disjoint sets $P$ of {\em places} and $T$ of {\em transitions}, a (total) {\em flow function} $f: P \times T \rightarrow \tau$, and an {\em initial marking} $M_0: P\longrightarrow  \{0,1\}$. 
A transition $t \in T$ can {\em fire} in a marking $M: P\longrightarrow  \{0,1\}$ if $\delta_\tau(M(p), f(p,t))$ is defined for all $p\in P$.
By firing, $t$ produces the marking $M' : P\longrightarrow  \{0,1\}$ where $M'(p)=\delta_\tau(M(p), f(p,t))$ for all $p\in P$, denoted by $M \edge{t} M'$.
The behavior of $\tau$-net $N$ is captured by a transition system $A_N$, called the {\em reachability graph} of $N$.
The states set $RS(N)$ of $A_N$ consists of all markings that can be reached from initial state $M_0$ by sequences of transition firings. 
The dependency number $d_p=\lvert \{ t \in T \mid f(p,t) \neq \nop \} \rvert $ of a place $p$ of $N$ is the number of transitions whose firing can possibly influence $p$ or be influenced by the marking of $p$. 
The \emph{dependency number} $d_N$ of a $\tau$-net $N$ is defined as $ d_N=\text{max}\{ d_p \mid p\in P\}$.
For $d\in \mathbb{N}$, a $\tau$-net is called {\em (dependency) $d$-restricted} if $d_N\leq d$.

\begin{example}
Figure~\ref{fig:type_net_rg} shows the type $\tau=\{\nop,\inp,\swap\}$ and the $2$-restricted $\tau$-net\\ $N=(\{R_1,R_2\}, \{a,b\}, f, M_0)$ with places $R_1, R_2$, flow-function $f(R_1,a)=f(R_2, b)=\inp$, $f(R_1,b)=\nop$, $f(R_2,a)=\swap$ and initial marking $M_0=(M_0(R_1),M_0(R_2))=(1,0)$.
Since $1\edge{\inp}0\in \tau$ and $0\edge{\swap}1\in \tau$, the transition $a$ can fire in $M_0$, which leads to the marking $M=(M(R_1), M(R_2))=(0,1)$.
After that, $b$ can fire, which results in the marking $M'=(M'(R_1), M'(R_2))=(0,0)$.
The reachability graph $A_N$ of $N$ is depicted on the right hand side of Figure~\ref{fig:type_net_rg}.
\end{example}

\textbf{$\tau$-Regions.}
Let $\tau\subseteq I$.
If an input $A$ of $\tau$-synthesis allows a positive decision, we want to construct a corresponding $\tau$-net $N$. 
TS represents the behavior of a modeled system by means of {\em global states} (states of TS) and transitions between them (events). 
Dealing with a Petri net, we operate with {\em local states} (places) and their changing (transitions), while the global states of a net are markings, i.e.,  combinations of local states.  
Since $A$ and $A_N$ must be isomorphic, $N$'s transitions correspond to $A$'s events.
The connection between global states in TS and local states in the sought net is given by 
{\em regions of TS} that mimic places: 
A $\tau$-region $R=(sup, sig)$ of $A=(S, E, \delta, \iota)$ consists of the \emph{support} $sup: S \rightarrow \{0,1\}$ and the \emph{signature} $sig: E\rightarrow E_\tau$ where every edge $s \edge{e} s'$ of $A$ leads to an edge $sup(s) \ledge{sig(e)} sup(s')$ of type $\tau$.
If $P=q_0\edge{e_1}\dots \edge{e_n}q_n$ is a path in $A$, then $P^R=sup(q_0)\ledge{sig(e_1)}\dots\ledge{sig(e_n)}sup(q_n)$ is a path in $\tau$.
We say $P^R$ is the \emph{image} of $P$ (under $R$).
Notice that $R$ is \emph{implicitly} defined by $sup(\iota)$ and $sig$:
Since $A$ is reachable, for every state $s\in S(A)$, there is a path $\iota\edge{e_1}\dots \edge{e_n}s_n$ such that $s=s_n$.
Thus, since $\tau$ is deterministic, we inductively obtain $sup(s_{i+1})$ by $sup(s_{i})\edge{e_i}sup(s_{i+1})$ for all $i\in \{0,\dots, n-1\}$ and $s_0 = \iota$.
Consequently, we can compute $sup$ and, thus, $R$ purely from $sup(\iota)$ and $sig$, cf. Figure~\ref{fig:image} and Example~\ref{ex:image}.
A region $(sup, sig)$ models a place $p$ and the associated part of the flow function $f$.
In particular, $f(p,e) = sig(e)$ and $M(p) = sup(s)$, for marking $M\in RS(N)$ that corresponds to $s\in S(A)$. 
Every set $\mathcal{R} $ of $\tau$-regions of $A$ defines the \emph{synthesized $\tau$-net} $N^{\mathcal{R}}_A=(\mathcal{R}, E, f, M_0)$ with  $f((sup, sig),e)=sig(e)$ and $M_0((sup, sig))=sup(\iota)$ for all $(sup, sig)\in \mathcal{R}, e\in E$.

\textbf{State and Event Separation.}
To ensure that the input behavior is captured by the synthesized net, we have to distinguish global states, and prevent the firings of transitions when their corresponding events are not present in TS. 
This is stated as so called {\em separation atoms} and {\em problems}. 
A pair $(s, s')$ of distinct states of $A$ defines a \emph{states separation atom} (SSP atom).
A $\tau$-region $R=(sup, sig)$ \emph{solves} $(s,s')$ if $sup(s)\not=sup(s')$.
If every SSP atom of $A$ is $\tau$-solvable then $A$ has the \emph{$\tau$-states separation property} ($\tau$-SSP, for short).
A pair $(e,s)$ of event $e\in E $ and state $s\in S$ where $e$ does not occur, that is $\neg s\edge{e}$, defines an \emph{event/state separation atom} (ESSP atom).
A $\tau$-region $R=(sup, sig)$ \emph{solves} $(e,s)$ if $sig(e)$ is not defined on $sup(s)$ in $\tau$, that is, $\neg sup(s)\edge{sig(e)}$.
If every ESSP atom of $A$ is $\tau$-solvable then $A$ has the \emph{$\tau$-event/state separation property} ($\tau$-ESSP, for short).
A set $\mathcal{R}$ of $\tau$-regions of $A$ is called $\tau$-\emph{admissible} if for each SSP and ESSP atom there is a $\tau$-region $R$ in $\mathcal{R}$ that solves it.
We say that $A$ is $\tau$-solvable if it has a $\tau$-admissible set.
The next lemma establishes the connection between the existence of $\tau$-admissible sets of $A$ and the existence of a $\tau$-net $N$ that solves $A$: 
\begin{lemma}[\cite{DBLP:series/txtcs/BadouelBD15}]\label{lem:admissible} 
A TS $A$ is isomorphic to the reachability graph of a $\tau$-net $N$ if and only if there is a $\tau$-admissible set $\mathcal{R}$ of $A$ such that $N=N^{\mathcal{R}}_A$.
\end{lemma}
\begin{figure}[t!]
\begin{minipage}{1.0\textwidth}
\centering
\begin{tikzpicture}[scale = 1.2]
\begin{scope}
\node (0) at (0,0) {\nscale{$0$}};
\node (1) at (1,0) {\nscale{$1$}};

\path (0) edge [->, out=-120,in=120,looseness=5] node[left, align =left] {\escale{$\nop$} \\ \escale{\free}} (0);
\path (1) edge [<-, out=60,in=-60,looseness=5] node[right, align=left] {\escale{$\nop$}  } (1);

\path (0) edge [<-] node[below] {\escale{$\inp$}} (1);
\node () at (0.5,-1) {\nscale{$ \tau_0 $}};

\end{scope}
\begin{scope}[xshift=3cm]
\node (0) at (0,-0.6) {\nscale{$s_0$}};
\node (1) at (0,0.6) {\nscale{$s_1$}};
\draw[-latex](-0.4,-0.6)to node[]{}(0);
\draw[-latex,out=60,in=-60](0)to node[right]{$a$}(1);
\draw[-latex,out=240,in=120](1)to node[left]{$a$}(0);
\node () at (0.0,-1) {\nscale{$ A_1$}};
\end{scope}
\begin{scope}[xshift=4.5cm]
\node (0) at (0,-0.6) {\nscale{$r_0$}};
\node (1) at (0,0.6) {\nscale{$r_1$}};
\draw[-latex](-0.4,-0.6)to node[]{}(0);
\draw[-latex,out=60,in=-60](0)to node[right]{$b$}(1);
\draw[-latex,out=240,in=120](1)to node[left]{$c$}(0);
\node () at (0.0,-1) {\nscale{$ A_2$}};
\end{scope}
\begin{scope}[xshift=6.5cm]
\node (0) at (0,0) {\nscale{$0$}};
\node (1) at (1.5,0) {\nscale{$1$}};

\path (0) edge [->, out=-120,in=120,looseness=5] node[left, align =left] {\escale{$\nop$} } (0);
\path (1) edge [<-, out=60,in=-60,looseness=5] node[right, align=left] {\escale{$\nop$} \\ \escale{\used} \\ \escale{\set} } (1);

\path (0) edge [<-, bend right= 30] node[below] {\escale{$ \swap$}\  } (1);
\path (0) edge [->, bend left= 30] node[above] {\escale{$\set,\swap$}} (1);

\node () at (0.75,-1) {\nscale{$\tau_1$}};
\end{scope}
\end{tikzpicture}
\caption{%
The type $ \tau_0 = \{ \nop, \inp, \free \}$, the TSs $A_1$ and $A_2$ and the type $ \tau_1 = \{ \nop, \swap, \used, \set \}$.}\label{fig:synthesis}
\end{minipage}

\begin{minipage}{\textwidth}
\begin{center}
\begin{tikzpicture}[new set = import nodes]
\tikzstyle{place}=[circle,thick,draw=blue!75,fill=blue!20,minimum size=6mm]
\tikzstyle{red place}=[place,draw=red!75,fill=red!20]
\tikzstyle{transition}=[rectangle,thick,draw=black!75,
  			  fill=black!20,minimum size=4mm]
			   \tikzstyle{every label}=[red]	
\begin{scope}[node distance=0.5cm,bend angle=45,auto]

\node [place] (R)[label=left: $R_1$, node distance =1.5cm]{};
\node [transition] (a)[right of =R, node distance =2cm]{$a$}
     	 edge [above] node {$\nscale{\swap}$}  (R);

	\node () at (0.75,-1.) {$N$};
\end{scope}

\begin{scope}[xshift=4.5cm,nodes={set=import nodes}]
		\node (A) at (0.8,-1.) {$A_N$};
		\node (0) at (0,0) {\nscale{(0)}};
		\node (1) at (2,0) {\nscale{(1)}};
		\draw[-latex](0,-0.6)to node[]{}(0);
		
\graph {(import nodes);
			0 ->[bend left =20, "\escale{$a$}"]1;
			1 ->[bend left =20, "\escale{$a$}"]0;
		};
\end{scope}
\end{tikzpicture}
\end{center}
\caption{%
The $1$-restricted $\tau_1$-net $N$, where $\tau_1$ is defined according to Figure~\ref{fig:synthesis} and $N=N_{A_1}^{\mathcal{R}}$ according to Example~\ref{ex:synthesis}, and its reachability graph $A_N$.}\label{fig:synthesized_net}
\end{minipage}
\begin{minipage}{1.0\textwidth}
\centering
\begin{tikzpicture}[scale = 1.2, new set = import nodes]
\begin{scope}[nodes={set=import nodes}]
\node () at (2.25,-0.75) {$A_3$};
\node (0) at (0,0) {\nscale{$\iota$}};
\foreach \i in {1,...,3} {\node (\i) at (\i*1.5cm,0) {\nscale{$s_\i$}}; } 
\draw[-latex](0,-0.4)to node[]{}(0);
\graph {
	(import nodes);
			0->["\escale{$a$}"]1->["\escale{$b$}"]2->["\escale{$c$}"]3;
			};
\end{scope}
\begin{scope}[xshift=6cm,nodes={set=import nodes}]
\node () at (2.25,-0.75) {$A_3^R$};
\foreach \i in {0,...,3} {\coordinate (\i) at (\i*1.5cm,0); } 
\foreach \i in {0,1,3} {\node (\i) at (\i) {\nscale{$1$}}; } 
\node (2) at (2) {\nscale{$0$}}; 
\draw[-latex](0,-0.4)to node[]{}(0);
\graph {
	(import nodes);
			0->["\escale{$\used$}"]1->["\escale{$\swap$}"]2->["\escale{$\set$}"]3;
			};
\end{scope}
\end{tikzpicture}
\end{minipage}
\caption{%
The TS $A_3$, a simple directed path and its image $A_3^R$ under $R$, with $R$ from Example~\ref{ex:image}.
}\label{fig:image}
\end{figure}
\begin{example}\label{ex:synthesis}
Let $\tau_0$, $\tau_1$, $A_1$ and $A_2$ be defined in accordance to Figure~\ref{fig:synthesis}.
The TS $A_1$ has no ESSP atoms.
Hence, it has the $\tau_0$-ESSP and $\tau_1$-ESSP.  
The only SSP atom of $A_1$ is $(s_0, s_1)$.
It is $\tau_1$-solvable by $R_1=(sup_1,sig_1)$ with $sup_1(s_0) = 0$, $sup_1(s_1) = 1$, $sig_1(a) = \swap$. 
Thus, $A_1$ has the $\tau_1$-admissible set $\mathcal{R}=\{R_1\}$, and the $\tau_1$-net $N=N_A^{\mathcal{R}}= (\{ R_1\}, \{ a\}, f, M_0)$ with $M_0(R_1) = sup_1(s_0)$ and $f(R_1,a) = sig_1(a)$ solves $A_1$. 
Figure~\ref{fig:synthesized_net} depicts $N$ (left) and its reachability graph $A_N$ (right).
The SSP atom $(s_0,s_1)$ is not $\tau_0$-solvable, thus, neither is $A_1$.
TS $A_2$ has ESSP atoms $(b,r_1)$ and $(c,r_0)$, which are both $\tau_1$-unsolvable. 
The only SSP atom $(r_0,r_1)$ in $A_2$ can be solved by the $\tau_1$-region $R_2=(sup_2, sig_2)$ with $sup_2(r_0) = 0$, $sup_2(r_1) = 1$, $sig_2(b) = \set$, $sig_2(c) = \swap$. 
Thus, $A_2$ has the $\tau_1$-SSP, but not the $\tau_1$-ESSP. 
None of the (E)SSP atoms of $A_2$ can be solved by any $\tau_0$-region. 
Notice that the $\tau_1$-region $R_2$ maps two events to a signature different from $\nop$.
Thus, in case of $d$-restricted $\tau_1$-synthesis, $R_2$ would not be valid for $d=1$. 
\end{example}

\begin{example}\label{ex:image}
Let $A_3$ be defined in accordance to Figure~\ref{fig:image} and $\tau_1$ according to Figure~\ref{fig:synthesis}.
It defines $sup(\iota)=1$, $sig(a)=\used$, $sig(b)=\swap$ and $sig(c)=\set$ implicitly a $\tau_1$-region $R=(sup, sig)$ of $A_3$ as follows:
$sup(s_1)=\delta_{\tau_1}(1,\used)=1$, $sup(s_2)=\delta_{\tau_1}(1,\swap)=0$ and $sup(s_3)=\delta_{\tau_1}(0,\set)=1$.
The image $A_3^R$ of $A_3$ (under R) is depicted on the right hand side of Figure~\ref{fig:image}.
One easily verifies that $\delta_{A_3}(s,e)=s'$ implies $\delta_{\tau_1}(sup(s), sig(e))=sup(s')$, cf. Figure~\ref{fig:synthesis}.
\end{example}

By Lemma~\ref{lem:admissible}, every $\tau$-admissible set $\mathcal{R}$ implies that $N^{\mathcal{R}}_A$ $\tau$-solves $A$. 
In this paper, we investigate the complexity of synthesising a solving $\tau$-net $N$ whose dependency number $d_N$ does not exceed a natural number $d$. 
Recall that if $\mathcal{R}$ is a set of $A$'s regions, then $\mathcal{R}$'s regions model places of the synthesized net $N^{\mathcal{R}}_A$.
Thus, $d_{N^{\mathcal{R}}_A} \leq d$ if and only if $\mathcal{R}$ is \emph{d-restricted}, that is, every region $R=(sup, sig)$ of $\mathcal{R}$ is \emph{$d$-restricted}: $\lvert \{e \in E \mid sig(e) \neq \nop \} \rvert \le d$.
By Lemma~\ref{lem:admissible}, this implies that there is a $d$-restricted $\tau$-net $N$ if and only if there is a $d$-restricted $\tau$-admissible set $\mathcal{R}$.
This finally leads to the following parameterized problem that is the main subject of this paper: 

\noindent\textbf{Dependency Restricted $\tau$-Synthesis (DR$\tau$S)}
\vspace*{-0.2cm}
\setlist[description]{font=\normalfont\itshape\space}
\begin{description}
	\item[Input: ] 
	\hspace*{0.85cm}a finite, reachable TS $A$, a natural number $d$.  
	\item[Parameter: ] $d$
	\item[Decide: ] 
	\hspace*{0.65cm}whether there exists a $d$-restricted $\tau$-admissible set $\mathcal{R}$ of $A$.
\end{description}

\section{Dependency $d$-Restricted $\tau$-Synthesis}\label{sct:dnr}%
%

For a start, we observe that, similar to (unrestricted) $\tau$-synthesis~\cite{DBLP:journals/corr/abs-1911-05834}, DR$\tau$S is in NP.
Moreover, there is a trivial reduction from $\tau$-synthesis to DR$\tau$S:
Since a $\tau$-region can map at most all events of a TS $A=(S,E,\delta,\iota)$ not to \nop, $A$ is $\tau$-solvable if and only if $A$ is $\tau$-solvable by $\vert E\vert $-restricted $\tau$-regions.
Thus, if $\tau$-synthesis is NP-complete, then DR$\tau$S is also NP-complete.

Let's argue that DR$\tau$S belongs to the complexity class XP.
Let $A=(S,E,\delta,\iota)$ be a TS,  $d\in \mathbb{N}$ and let $\vert A\vert$ be the maximum number of edges that $A$ possibly has, that is,   $\vert A\vert=\vert S\vert^2 \vert E\vert $.
A $\tau$-region $R=(sup, sig)$ is implicitly defined by $sup(\iota)$ and $sig$.
We are interested in regions of $A$ for which there is an $i\in \{0,\dots, d\}$ such that $\vert \{e\in E\mid sig(e)\not=\nop\}\vert=i$.
For every event $e\in E$, we have at most $\vert \tau\vert -1\leq 7$ interactions that are different from $\nop$.
Since $sup(\iota)\in \{0,1\}$, we have to consider at most $2\cdot 7^d \cdot \sum_{i=0}^{d}\binom{\vert E\vert}{i}$ regions at all, which can be estimated by $\mathcal{O}(7^d \vert A\vert^d )$

To check if the chosen signature actually implies regions of $A$ and to solve the (E)SSP atoms of $A$, we need to construct the regions explicitly, that is, we have to compute $sup$.
To do so, we firstly compute a spanning tree $A'$ of $A$, which is doable in time $\mathcal{O}(\vert A\vert^2 )$ by the algorithm of Tarjan~\cite{DBLP:journals/networks/Tarjan77} and needs  to be done only once.
In $A'$, there is exactly one path from $\iota$ to $s$ for all $s\in S$, and $A'$ has $\vert S\vert -1$ edges.
Thus, having a spanning tree, $sup(\iota)$ and $sig$, it costs time at most $\mathcal{O}(\vert A\vert) $ to compute $sup$. 
The effort to compute all \emph{potentially} interesting regions explicitly is thus at most $\mathcal{O}(7^d \vert A\vert^{d+1})$.
After that, we check for any fixed potential region if it is actually a well-defined region, that is, whether $s\edge{e}s'$ implies $sup(s)\ledge{sig(e)}sup(s')$.
For a fixed region, this is doable in time $\mathcal{O}( \vert A\vert )$.
Thus the effort to compute all interesting regions of $A$ is $\mathcal{O}(7^d \vert A\vert^{d+2})$.

For a fixed separation atom $(s,s')$ or $(e,s)$ we simply have to check if $sup(s)\not=sup(s')$ or if $\delta_\tau(sup(s), sig(e))$ is not defined, respectively, which is doable in time $\mathcal{O}(\vert A\vert)$. 
Since we have at most $\mathcal{O}(\vert A\vert^2)$ separation atoms and at most $\mathcal{O}(7^d \vert A\vert^d )$ regions, the check for the (E)SSP is doable in time $\mathcal{O}(7^d\vert A\vert^{d+2})$.
Finally, if we add up the effort to get all interesting regions and the effort to check whether these regions witness the (E)SSP of $A$, then we obtain that the effort of the problem is bounded by $\mathcal{O}(7^d\vert A\vert^{d+2})$.

On the other hand, in this section, we argue that DR$\tau$S is $W[2]$-hard for a range of Boolean types.
The following theorem presents the result through the enumeration of these types. 
\begin{theorem}\label{the:w2_result}
Dependency $d$-Restricted $\tau$-Synthesis is $W[2]$-hard if 
\begin{enumerate}
\item\label{the:w2_result_nop_inp_set}
$\tau\supseteq \{\nop,\inp,\set\}$ or $\tau\supseteq \{\nop,\out,\res\}$,
\item\label{the:w2_result_nop_set_res}
$\tau=\{\nop,\set,\res\}\cup\omega$ or $\tau=\{\nop,\set,\res,\swap\}\cup\omega$ and $\emptyset\not=\omega\subseteq\{\free,\used\}$,%
\item\label{the:w2_result_nop_set_swap}
$\tau=\{\nop,\set,\swap\}\cup\omega$, $\tau=\{\nop, \out, \set,\swap\}\cup\omega$, $\tau=\{\nop,\res,\swap\}\cup\omega$ or \\ $\tau=\{\nop, \inp, \res,\swap \}\cup\omega$ and $\emptyset\not=\omega\subseteq\{\free,\used\}$,
\item\label{the:w2_result_nop_inp_res_swap}
$\tau=\{\nop, \inp, \res,\swap\}$ or $\tau=\{\nop, \out, \set, \swap\}$, 
%
%
\end{enumerate}
\end{theorem}

Notice that, by the discussion above, for the types of Theorem~\ref{the:w2_result}, NP-completeness of DR$\tau$S follows by the NP-completeness of $\tau$-synthesis~\cite[p.~3]{DBLP:journals/corr/abs-1911-05834}.
The proof of Theorem~\ref{the:w2_result} bases on parameterized reductions of the problem \emph{Hitting Set}, which is known to be $W[2]$-complete~(see e.g.~\cite{DBLP:books/sp/CyganFKLMPPS15}).
The problem \emph{Hitting Set} is defined as follows: 

\noindent\textbf{Hitting Set (HS)}
\vspace*{-0.2cm}
\setlist[description]{font=\normalfont\itshape\space}
\begin{description}
	\item[Input: ] 
	\hspace*{0.75cm} a finite set $\mathfrak{U}$, a set $M=\{M_1,\dots, M_{m}\}$ of subsets of $\mathfrak{U}$ with $M_i=\{X_{i_1},\dots, X_{i_{m_i}}\}$ \\ 
	\hspace*{1.45cm}and $i_1 < \dots < i_{m_i}$ for all $i\in \{1,\dots,m\}$, a natural number $\kappa$.
	\item[Parameter: ] $\kappa$
	\item[Decide: ] 
	\hspace*{0.65cm}whether there is a set $S\subseteq \mathfrak{U}$ such that $\vert S\vert \leq \kappa$ and $S\cap M_i\not=\emptyset$ for every $i\in \{1,\dots, m\}$.
	\end{description}

\textbf{The General Reduction Idea}.
An input $I=(\mathfrak{U}, M, \kappa)$ of HS, where $M=\{M_1,\dots, M_m\}$, is reduced to an instance $(A^\tau_I, d)$ of DR$\tau$S with TS $A^\tau_I$ and $d=f(\kappa)$, for some linear function $f$. 
For every $i\in \{1,\dots, m\}$, the TS $A^\tau_I$ has a directed labeled path 
\begin{center}
\begin{tikzpicture}[new set = import nodes]
\begin{scope}[nodes={set=import nodes}]%
		\node (top) at (-1,0) {$P_i=$};
		\foreach \i in {0,...,5} { \coordinate (\i) at (\i*1.8cm,0) ;}
		 \node (0) at (0) {\nscale{$s_{i,0}$}};
		 \node (dots) at (1) {$\dots$};
		 \node (2) at (2) {\nscale{$s_{i,i_{\ell-1}}$}};
		 \node (3) at (3) {\nscale{$s_{i,i_{\ell}}$}};
		 \node (dots2) at (4) {$\dots$};
		 \node (5) at (5) {\nscale{$s_{i,i_{m_i}}$}};
		\graph {
	(import nodes);
			0->["\escale{$X_{i_1}$}"]dots->["\escale{$X_{i_{\ell-1}}$}"]2->["\escale{$X_{i_{\ell}}$}"]3->["\escale{$X_{i_{\ell+1}}$}"]dots2->["\escale{$X_{i_{m_i}}$}"]5;
			};
\end{scope}
\end{tikzpicture}
\end{center}
that represents the set $M_i=\{X_{i_1},\dots, X_{i_{m_i}}\}$ and uses its elements as events. 
The TS $A^\tau_I$ is then composed in such a way that for some ESSP atom $\alpha$ of $A^\tau_I$ the following is satisfied: 
If $R=(sup, sig)$ is a $d$-restricted $\tau$-region that solves $\alpha$, then $sup(s_{i,0})\not=sup(s_{i,i_{m_i}})$ for all $i\in \{1,\dots, m\}$.
Since the image $P_i^R$ of $P_i$ is a directed path in $\tau$, by $sup(s_{i,0})\not=sup(s_{i,i_{m_i}})$, there has to be an element $X\in M_i$ such that $s\edge{X}s'\in P_i$ implies $sup(s)\not=sup(s')$.
That is, the image $sig(X)$ of $X$ causes a state change on $P_i^R$ in $\tau$.
In particular, this implies $sig(X)\not=\nop$.
The following visualisation of $P_i^R$ sketches the situation for a region $R=(sup, sig)$, where $sup(s_{i,0})=\dots =sup(s_{i,i_{\ell-1}})=0$ and $sup(s_{i,i_\ell})=\dots =sup(s_{i,i_{m_i}})=1$ and $sig(X_{i_\ell})=\set$ and $sig(X_{i_k})=\nop$ for all $k\in \{1,\dots, m_i\}\setminus\{\ell\}$:
\begin{center}
\begin{tikzpicture}[new set = import nodes]
\begin{scope}[nodes={set=import nodes}]%
		\node (top) at (-1.2,0) {$P_i^R=$};
		\foreach \i in {0,...,5} { \coordinate (\i) at (\i*2.6cm,0) ;}
		\foreach \i in {3} {\fill[green!20, rounded corners] (\i) +(-0.7,-0.35) rectangle +(6,0.7);}
		 \node (0) at (0) {\nscale{$sup(s_{i,0})$}};
		 \node (dots) at (1) {$\dots$};
		 \node (2) at (2) {\nscale{$sup(s_{i,i_{\ell-1}})$}};
		 \node (3) at (3) {\nscale{$sup(s_{i,i_{\ell}})$}};
		 \node (dots2) at (4) {$\dots$};
		 \node (5) at (5) {\nscale{$sup(s_{i,i_{m_i}})$}};
		\graph {
	(import nodes);
			0->["\escale{$sig(X_{i_1})$}"]dots->["\escale{$sig(X_{i_{\ell-1}})$}"]2->["\escale{$sig(X_{i_{\ell}})$}"]3->["\escale{$sig(X_{i_{\ell+1}})$}"]dots2->["\escale{$sig(X_{i_{m_i}})$}"]5;
			};
\end{scope}
\begin{scope}[yshift=-0.5cm,nodes={set=import nodes}]%
		\foreach \i in {0,...,5} { \coordinate (\i) at (\i*2.6cm,0) ;}
		\foreach \i in {0,...,5} { \coordinate (m\i) at (\i*2.6cm+1.3cm,0) ;}
		\foreach \i in {0,2} {  \node (\i) at (\i) {\nscale{$0$}};}
		\foreach \i in {3,5} {  \node (\i) at (\i) {\nscale{$1$}};}
		\foreach \i in {m0,m1,m3,m4} {  \node (\i) at (\i) {\nscale{$\nop$}};}
		 \node (m2) at (m2) {\nscale{$\set$}};		
\end{scope}
\end{tikzpicture}
\end{center}
It is simultaneously true for all paths $P_1,\dots, P_m$ representing the sets $M_1,\dots, M_m$, that on each path there is a (not necessarily unique) $X$ satisfying $sig(X)\not=\nop$.
Moreover, the reduction ensures that $\vert \{X\in \mathfrak{U}\mid sig(X)\not=\nop\}\vert \leq \kappa$.
In other words, $S=\{X\in \mathfrak{U}\mid sig(X)\not=\nop\}$ defines a sought hitting set of $I$. 
Thus, if $(A^\tau_I, d)$ is a yes-instance of DR$\tau$S, implying the solvability of $\alpha$, then $I=(\mathfrak{U}, M, \kappa)$ is a yes-instance of HS.

Conversely, if $I=(\mathfrak{U}, M, \kappa)$ is a yes-instance, then there is a fitting $\tau$-region of $A^\tau_I$ that solves $\alpha$.
The reduction ensures that the $d$-restricted $\tau$-solvability of $\alpha$ implies that all (E)SSP atoms of $A^\tau_I$ are solvable by $d$-restricted $\tau$-regions.
Thus, $(A^\tau_I,d)$ is a yes-instance, too.

In the following, we present the corresponding reductions and show that the solvability of $\alpha$ implies the existence of a sought-for hitting set. 
Moreover, we argue that the existence of a sought set implies the $\tau$-solvability of $\alpha$ and, finally, the $\tau$-solvability of $A^\tau_I$.

As an instance, the following (running) example serves for all concrete reductions that we present, to simplify the understanding of the reductions' formal descriptions. 
\begin{example}\label{ex:hitting_set}
The input $I=(\mathfrak{U}, M, \kappa)$ is defined by $\mathfrak{U}=\{X_1,X_2,X_3,X_4\}$ and $M=\{M_1,M_2,M_3,M_4\}$, where $M_1=\{X_1,X_2\}$, $M_2=\{X_2,X_3\}$, $M_3=\{X_1,X_4\}$ and $M_4=\{X_1,X_3,X_4\}$, and $\kappa=2$.
A fitting hitting set of $M$ is given by $S=\{X_1,X_3\}$.
\end{example}

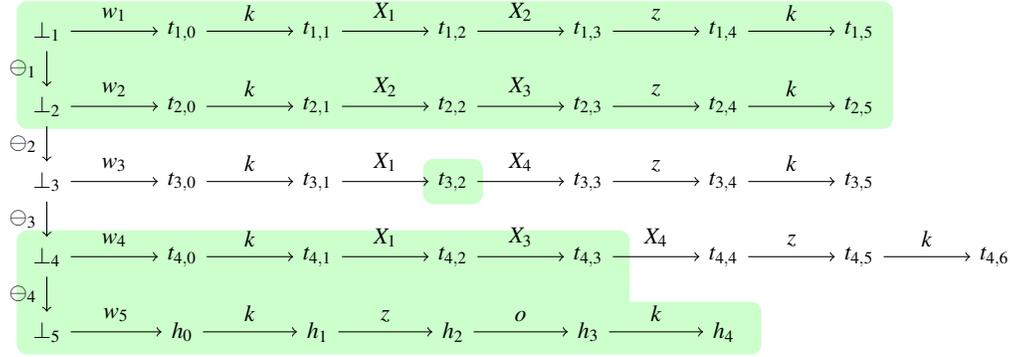
\begin{figure}[t!]
\begin{center}
\begin{tikzpicture}[new set = import nodes]
\begin{scope}[nodes={set=import nodes}]
		\coordinate (bot1) at (0,0);
		\foreach \i in {bot1} {\fill[green!20, rounded corners] (\i) +(-0.4,-1.3) rectangle +(11.25,0.4);}
		\foreach \i in {0,...,5} {\coordinate (\i) at (\i*1.8cm+1.8cm,0);}
		\node (bot1) at (bot1) {\nscale{$\bot_1$}};
		\foreach \i in {0,...,5} {\node (\i) at (\i) {\nscale{$t_{1,\i}$}};}
		\graph {
	(import nodes);
			bot1->["\escale{$w_1$}"]0->["\escale{$k$}"]1->["\escale{$X_1$}"]2->["\escale{$X_2$}"]3->["\escale{$z$}"]4->["\escale{$k$}"]5;
			};
\end{scope}
\begin{scope}[yshift=-1cm, nodes={set=import nodes}]
		\coordinate (bot2) at (0,0);
		\foreach \i in {0,...,5} {\coordinate (\i) at (\i*1.8cm+1.8cm,0);}
		\node (bot2) at (bot2) {\nscale{$\bot_2$}};
		\foreach \i in {0,...,5} {\node (\i) at (\i) {\nscale{$t_{2,\i}$}};}			
		\graph {
	(import nodes);
			bot2->["\escale{$w_2$}"]0->["\escale{$k$}"]1->["\escale{$X_2$}"]2->["\escale{$X_3$}"]3->["\escale{$z$}"]4->["\escale{$k$}"]5;
			};
\end{scope}
\begin{scope}[yshift=-2cm, nodes={set=import nodes}]
		\coordinate (bot3) at (0,0);
		\foreach \i in {0,...,5} {\coordinate (\i) at (\i*1.8cm+1.8cm,0);}
		\foreach \i in {2} {\fill[green!20, rounded corners] (\i) +(-0.4,-0.3) rectangle +(0.4,0.3);}
		\node (bot3) at (bot3) {\nscale{$\bot_3$}};
		\foreach \i in {0,...,5} {\node (\i) at (\i) {\nscale{$t_{3,\i}$}};}				
		\graph {
	(import nodes);
			bot3->["\escale{$w_3$}"]0->["\escale{$k$}"]1->["\escale{$X_1$}"]2->["\escale{$X_4$}"]3->["\escale{$z$}"]4->["\escale{$k$}"]5;
			};
\end{scope}
\begin{scope}[yshift=-3cm, nodes={set=import nodes}]%
		\coordinate (bot4) at (0,0);
		\foreach \i in {bot4} {\fill[green!20, rounded corners] (\i) +(-0.4,-0.8) rectangle +(7.75,0.35);}
		\foreach \i in {0,...,6} {\coordinate (\i) at (\i*1.8cm+1.8cm,0);}
		\node (bot4) at (bot4) {\nscale{$\bot_4$}};
		\foreach \i in {0,...,6} {\node (\i) at (\i) {\nscale{$t_{4,\i}$}};}					
		\graph {
	(import nodes);
			bot4->["\escale{$w_4$}"]0->["\escale{$k$}"]1->["\escale{$X_1$}"]2->["\escale{$X_3$}"]3->["\escale{$X_4$}"]4->["\escale{$z$}"]5->["\escale{$k$}"]6;
			};
\end{scope}
\begin{scope}[yshift=-4cm,nodes={set=import nodes}]%
		\coordinate (bot5) at (0,0);
		\foreach \i in {bot5} {\fill[green!20, rounded corners] (\i) +(-0.4,-0.3) rectangle +(9.5,0.4);}
		\node (bot5) at (0,0) {\nscale{$\bot_{5}$}};
		\foreach \i in {0,...,4} { \coordinate (\i) at (1.8cm+ \i*1.8cm,0) ;}
		\foreach \i in {0,...,4} { \node (\i) at (\i) {\nscale{$h_{\i}$}};}
		\graph {
	(import nodes);
			bot5->["\escale{$w_{5}$}"]0->["\escale{$k$}"]1->["\escale{$z$}"]2->["\escale{$o$}"]3->["\escale{$k$}"]4;
			};
\end{scope}

\path (bot1) edge [->] node[left] {\nscale{$\ominus_1$} } (bot2);
\path (bot2) edge [->] node[left] {\nscale{$\ominus_2$} } (bot3);
\path (bot3) edge [->] node[left] {\nscale{$\ominus_3$} } (bot4);
\path (bot4) edge [->] node[left] {\nscale{$\ominus_4$} } (bot5);
\end{tikzpicture}
\end{center}
\caption{The TS $A^\tau_I$, where $\tau\supseteq \{\nop,\inp,\set\}$ and $I$ originates from Example~\ref{ex:hitting_set}.
The green colored area sketches the states that are mapped to $1$ by the region $R^{X,2}_{3,2}$ solving $(X_4, s)$ for all $s\in \{\bot_3,t_{3,0},t_{3,1}\}$.}\label{fig:nop_inp_set}
\end{figure}
%
\subsection{The Proof of Theorem~\ref{the:w2_result}.\ref{the:w2_result_nop_inp_set}}

\textbf{Theorem~\ref{the:w2_result}.\ref{the:w2_result_nop_inp_set}: The Reduction.}
In accordance to our general approach, we first define $d=\kappa+2$.
Next, we introduce the TS $A^\tau_I$.
Figure~\ref{fig:nop_inp_set} provides a concrete example of $A^\tau_I$, where $I$ corresponds to Example~\ref{ex:hitting_set}.
The TS $A^\tau_I$ has the following gadget $H$ that applies the events $k, z$ and $o$ and provides the atom $\alpha=(k,h_2)$:
\begin{center}
\begin{tikzpicture}[new set = import nodes]
\begin{scope}[nodes={set=import nodes}]%
		\node (top) at (-1.8,0) {\nscale{$\bot_{m+1}$}};
		\foreach \i in {0,...,4} { \coordinate (\i) at (\i*1.8cm,0) ;}
		\foreach \i in {0,...,4} { \node (\i) at (\i) {\nscale{$h_{\i}$}};}
		\graph {
	(import nodes);
			top->["\escale{$w_{m+1}$}"]0->["\escale{$k$}"]1->["\escale{$z$}"]2->["\escale{$o$}"]3->["\escale{$k$}"]4;
			};
\end{scope}
\end{tikzpicture}
\end{center}
For all $i\in \{1,\dots, m\}$, the TS $A^\tau_I$ has the following gadget $T_i$ that applies $w_i,k,z$ and the elements of $M_i=\{X_{i_1},\dots, X_{i_{m_i}}\}$ as events:
\begin{center}
\begin{tikzpicture}[new set = import nodes]
\begin{scope}[nodes={set=import nodes}]%
		\coordinate (0) at (0,0);
		\coordinate (1) at (2,0);
		\coordinate (2) at (4,0);
		\coordinate (dots) at (6,0);
		\coordinate (3) at (8,0);
		\coordinate (4) at (10.25,0);
		\coordinate (5) at (12.5,0);
		
		\node (0) at (0) {\nscale{$\bot_i$}};
		\node (1) at (1) {\nscale{$t_{i,0}$}};	
		\node (2) at (2) {\nscale{$t_{i,1}$}};			
		\node (dots) at (dots) {$\dots$};
		\node (3) at (3) {\nscale{$t_{i,m_i+1}$}};
		\node (4) at (4) {\nscale{$t_{i,m_i+2}$}};
		\node (5) at (5) {\nscale{$t_{i,m_i+3}$}};					
		\graph {
	(import nodes);
			0 ->["\escale{$w_i$}"]1->["\escale{$k$}"]2->["\escale{$X_{i_1}$}"]dots->["\escale{$X_{i_{m_i}}$}"]3->["\escale{$z$}"]4->["\escale{$k$}"]5;
			};
\end{scope}
\end{tikzpicture}
\end{center}
The TS $A^\tau_I$ has the events $\ominus_1,\dots, \ominus_m$ to connect the gadgets $T_1,\dots, T_m$ and $H$ by $\bot_1\edge{\ominus_1}\dots\edge{\ominus_{m}}\bot_{m+1}$.
The initial state of $A^\tau_I$ is $\bot_1$.

\textbf{Theorem~\ref{the:w2_result}.\ref{the:w2_result_nop_inp_set}: The Solvability of $\alpha$ Implies a Hitting Set.}
We argue for $\tau\supseteq\{\nop,\inp,\set\}$, the hardness of the other types follows by symmetry.
In the following, we argue that if there is a $d$-restricted $\tau$-region $R=(sup, sig)$ that solves $\alpha$, then $I$ has a hitting set of size at most $\kappa$.
Let $R=(sup, sig)$ be such a $\tau$-region.
Since $R$ solves $\alpha$, we have either $sig(k)\in \{\inp, \used\}$ and $sup(h_2)=0$ or $sig(k)\in \{\out, \free\}$ and $sup(h_2)=1$.
In what follows, we consider to the former case. 
The proof for the latter case is symmetrical. 

If $sig(k)=\inp$ and $sup(h_2)=0$, then $s\edge{k}s'$ implies $sup(s)=1$ and $sup(s')=0$.
By $sup(h_2)=0$ and $sup(h_3)=1$, we get $sig(o)\in \{\out,\set,\swap\}$.
In particular, since $R$ is $d$-restricted, there are at most $\kappa$ events left that have a signature different from \nop.
By $sup(h_1)=sup(h_2)=0$ and $h_1\edge{z}h_2$, we have $sig(z)\in \{\nop,\res,\free\}$.
Moreover, by $sup(t_{i,m_i+2})=1$ and $\edge{z}t_{i,m_i+2}$, we have $sig(z)=\nop$.
By $sig(k)=\inp$ and $sig(z)=\nop$, we conclude $sup(t_{i,1})=0$ and $sup(t_{i,m_i+1})=1$ for all $i\in \{1,\dots, m\}$.
Consequently, for every $i\in \{1,\dots, m\}$, there is $X\in M_i$ such that $sig(X)\in \{\out,\set,\swap\}$.
Otherwise a state change from $0$ to $1$ would not be possible.
Since $R$ is $d$-restricted and $sig(k) \neq \nop \neq sig(o)$, we get $\vert \{X\in \mathfrak{U} \mid sig(X)\not=\nop \}\vert \leq \kappa$.
This implies that $S=\{X\in \mathfrak{U} \mid sig(X)\not=\nop\}$ is a fitting hitting set of $I$.

If $sig(k)=\used$ and $sup(h_2)=0$, then $s\edge{k}s'$ implies $sup(s)=sup(s')=1$.
By $sup(h_1)=sup(h_3)=1$ and $sup(h_2)=0$, we get $sig(z)\in \{\inp,\res,\swap\}$ and $sig(o)\in \{\out,\set,\swap\}$.
By $sup(t_{i,m_i+2})=1$ and $\edge{z}t_{i,m_i+2}$, we get $sig(z)=\swap$. 
Since $R$ is $d$-restricted, there are at most $\kappa-1$ events left whose signature is different from \nop.
Moreover, by $sig(k)=\used$ and $sig(z)=\swap$, we have $sup(t_{i,1})=1$ and $sup(t_{i,m_i+1})=0$ for all $i\in \{1,\dots, m\}$.
Just like before, we conclude that $S=\{X\in \mathfrak{U} \mid sig(X)\not=\nop\}$ is a sought hitting set of $I$.

Conversely, a $\kappa$-HS of $(\mathfrak{U}, M,\kappa)$ implies the $\tau$-solvability of $A^\tau_I$, which is the statement of the following lemma.
Due to space restrictions, we omit the proof which can be found in \cite{DBLP:journals/corr/abs-2007-12372}.
\begin{lemma}\label{lem:w2_result_nop_inp_set}
Let $\tau$ be a type of nets in correspondence of Theorem~\ref{the:w2_result}.\ref{the:w2_result_nop_inp_set}.
If $(\mathfrak{U}, M,\kappa)$ has a $\kappa$-HS, then there is a $d$-restricted admissible set of $A_\tau^I$.
\end{lemma}

\subsection{The Proof of Theorem~\ref{the:w2_result}.\ref{the:w2_result_nop_set_res}}%

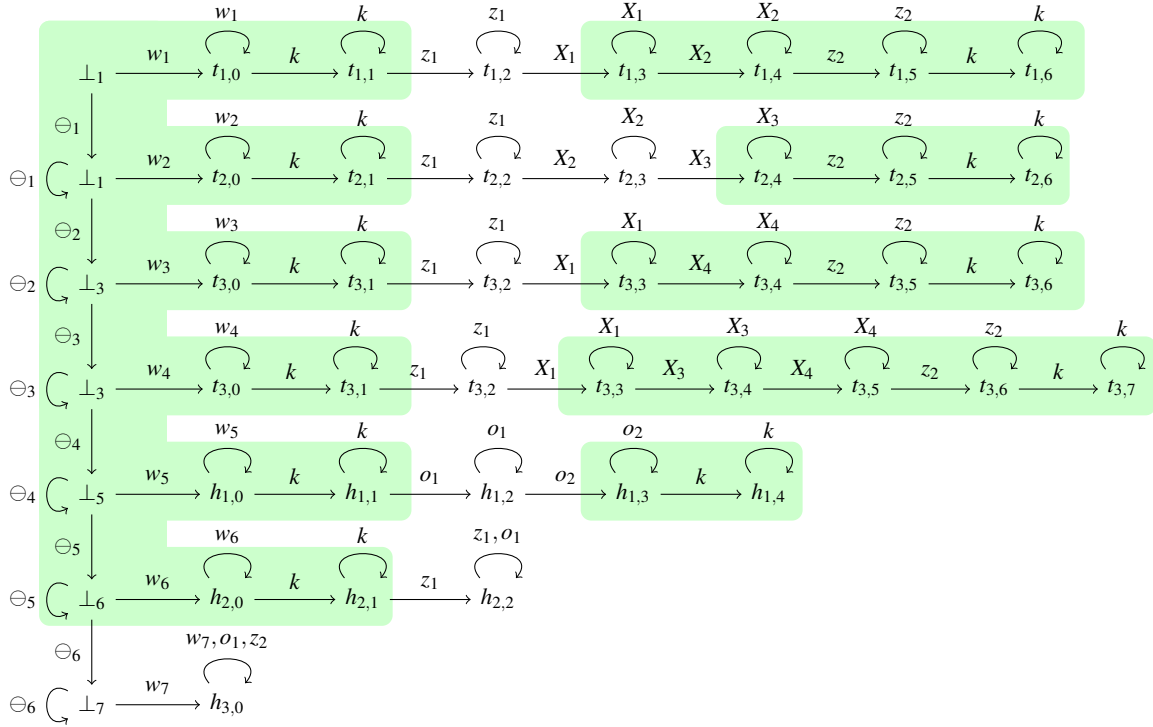
\begin{figure}[t!]
\begin{center}
\begin{tikzpicture}[new set = import nodes]
\begin{scope}[nodes={set=import nodes}]%
		\coordinate (bot1) at (0,0);
		\foreach \i in {0,...,6} {\coordinate (\i) at (\i*1.8+1.8,0);}
		\foreach \i in {bot1} {\fill[green!20, rounded corners] (\i) +(-0.7,-0.35) rectangle +(4.25,0.7);}
		\foreach \i in {bot1} {\fill[green!20, rounded corners] (\i) +(-0.7,-7) rectangle +(1,0.7);}
		\foreach \i in {3} {\fill[green!20, rounded corners] (\i) +(-0.7,-0.35) rectangle +(6,0.7);}
		\node (bot1) at (bot1) {\nscale{$\bot_1$}};
		\foreach \i in {0,...,6} { \node (\i) at (\i) {\nscale{$t_{1,\i}$}}; }
		\path (0) edge [->, out=130,in=50,looseness=3] node[above] {\nscale{$w_1$} } (0);					
		\path (1) edge [->, out=130,in=50,looseness=3] node[above] {\nscale{$k$} } (1);	
		\path (2) edge [->, out=130,in=50,looseness=3] node[above] {\nscale{$z_1$} } (2);	
		\path (3) edge [->, out=130,in=50,looseness=3] node[above] {\nscale{$X_1$} } (3);	
		\path (4) edge [->, out=130,in=50,looseness=3] node[above] {\nscale{$X_2$} } (4);
		\path (5) edge [->, out=130,in=50,looseness=3] node[above] {\nscale{$z_2$} } (5);
		\path (6) edge [->, out=130,in=50,looseness=3] node[above] {\nscale{$k$} } (6);				
		
		\graph {
	(import nodes);
			bot1->["\escale{$w_1$}"]0->["\escale{$k$}"]1->["\escale{$z_1$}"]2->["\escale{$X_1$}"]3->["\escale{$X_2$}"]4->["\escale{$z_2$}"]5->["\escale{$k$}"]6;
			};
\end{scope}
\begin{scope}[yshift=-1.4cm, nodes={set=import nodes}]%
		\coordinate (bot2) at (0,0);
		\foreach \i in {0,...,6} {\coordinate (\i) at (\i*1.8+1.8,0);}
		\foreach \i in {bot2} {\fill[green!20, rounded corners] (\i) +(-0.7,-0.35) rectangle +(4.25,0.7);}
		\foreach \i in {4} {\fill[green!20, rounded corners] (\i) +(-0.7,-0.35) rectangle +(4,0.7);}
		\node (bot2) at (bot2) {\nscale{$\bot_1$}};
		\foreach \i in {0,...,6} { \node (\i) at (\i) {\nscale{$t_{2,\i}$}}; }
		
		\path (0) edge [->, out=130,in=50,looseness=3] node[above] {\nscale{$w_2$} } (0);					
		\path (1) edge [->, out=130,in=50,looseness=3] node[above] {\nscale{$k$} } (1);	
		\path (2) edge [->, out=130,in=50,looseness=3] node[above] {\nscale{$z_1$} } (2);	
		\path (3) edge [->, out=130,in=50,looseness=3] node[above] {\nscale{$X_2$} } (3);	
		\path (4) edge [->, out=130,in=50,looseness=3] node[above] {\nscale{$X_3$} } (4);
		\path (5) edge [->, out=130,in=50,looseness=3] node[above] {\nscale{$z_2$} } (5);
		\path (6) edge [->, out=130,in=50,looseness=3] node[above] {\nscale{$k$} } (6);						
		\graph {
	(import nodes);
			bot2->["\escale{$w_2$}"]0->["\escale{$k$}"]1->["\escale{$z_1$}"]2->["\escale{$X_2$}"]3->["\escale{$X_3$}"]4->["\escale{$z_2$}"]5->["\escale{$k$}"]6;
			};
\end{scope}
\begin{scope}[yshift=-2.8cm, nodes={set=import nodes}]%
		\coordinate (bot3) at (0,0);
		\foreach \i in {0,...,6} {\coordinate (\i) at (\i*1.8+1.8,0);}
		\foreach \i in {bot3} {\fill[green!20, rounded corners] (\i) +(-0.7,-0.35) rectangle +(4.25,0.7);}
		\foreach \i in {3} {\fill[green!20, rounded corners] (\i) +(-0.7,-0.35) rectangle +(6,0.7);}
		\node (bot3) at (bot3) {\nscale{$\bot_3$}};
		\foreach \i in {0,...,6} { \node (\i) at (\i) {\nscale{$t_{3,\i}$}}; }
		
		\path (0) edge [->, out=130,in=50,looseness=3] node[above] {\nscale{$w_3$} } (0);					
		\path (1) edge [->, out=130,in=50,looseness=3] node[above] {\nscale{$k$} } (1);	
		\path (2) edge [->, out=130,in=50,looseness=3] node[above] {\nscale{$z_1$} } (2);	
		\path (3) edge [->, out=130,in=50,looseness=3] node[above] {\nscale{$X_1$} } (3);	
		\path (4) edge [->, out=130,in=50,looseness=3] node[above] {\nscale{$X_4$} } (4);
		\path (5) edge [->, out=130,in=50,looseness=3] node[above] {\nscale{$z_2$} } (5);
		\path (6) edge [->, out=130,in=50,looseness=3] node[above] {\nscale{$k$} } (6);			
		\graph {
	(import nodes);
			bot3->["\escale{$w_3$}"]0->["\escale{$k$}"]1->["\escale{$z_1$}"]2->["\escale{$X_1$}"]3->["\escale{$X_4$}"]4->["\escale{$z_2$}"]5->["\escale{$k$}"]6;
			};
\end{scope}
\begin{scope}[yshift=-4.2cm, nodes={set=import nodes}]%
		\coordinate (bot4) at (0,0);
		\foreach \i in {0,...,7} {\coordinate (\i) at (\i*1.7+1.8,0);}
		\foreach \i in {bot4} {\fill[green!20, rounded corners] (\i) +(-0.7,-0.35) rectangle +(4.25,0.7);}
		\foreach \i in {3} {\fill[green!20, rounded corners] (\i) +(-0.7,-0.35) rectangle +(7.25,0.7);}
		\node (bot4) at (bot4) {\nscale{$\bot_3$}};
		\foreach \i in {0,...,7} { \node (\i) at (\i) {\nscale{$t_{3,\i}$}}; }
		
		\path (0) edge [->, out=130,in=50,looseness=3] node[above] {\nscale{$w_4$} } (0);					
		\path (1) edge [->, out=130,in=50,looseness=3] node[above] {\nscale{$k$} } (1);	
		\path (2) edge [->, out=130,in=50,looseness=3] node[above] {\nscale{$z_1$} } (2);	
		\path (3) edge [->, out=130,in=50,looseness=3] node[above] {\nscale{$X_1$} } (3);	
		\path (4) edge [->, out=130,in=50,looseness=3] node[above] {\nscale{$X_3$} } (4);
		\path (5) edge [->, out=130,in=50,looseness=3] node[above] {\nscale{$X_4$} } (5);
		\path (6) edge [->, out=130,in=50,looseness=3] node[above] {\nscale{$z_2$} } (6);
		\path (7) edge [->, out=130,in=50,looseness=3] node[above] {\nscale{$k$} } (7);					
		\graph {
	(import nodes);
			bot4->["\escale{$w_4$}"]0->["\escale{$k$}"]1->["\escale{$z_1$}"]2->["\escale{$X_1$}"]3->["\escale{$X_3$}"]4->["\escale{$X_4$}"]5->["\escale{$z_2$}"]6->["\escale{$k$}"]7;
			};
\end{scope}
\begin{scope}[yshift=-5.6cm,nodes={set=import nodes}]%
		\coordinate (bot5) at (0,0);
		\foreach \i in {0,...,4} { \coordinate (\i) at (1.8cm+ \i*1.8cm,0) ;}
		\foreach \i in {bot5} {\fill[green!20, rounded corners] (\i) +(-0.7,-0.35) rectangle +(4.25,0.7);}
		\foreach \i in {3} {\fill[green!20, rounded corners] (\i) +(-0.7,-0.35) rectangle +(2.25,0.7);}
		\node (bot5) at (bot5) {\nscale{$\bot_{5}$}};
		\foreach \i in {0,...,4} { \node (\i) at (\i) {\nscale{$h_{1,\i}$}};}
		
		\path (0) edge [->, out=130,in=50,looseness=3] node[above] {\nscale{$w_5$} } (0);					
		\path (1) edge [->, out=130,in=50,looseness=3] node[above] {\nscale{$k$} } (1);	
		\path (2) edge [->, out=130,in=50,looseness=3] node[above] {\nscale{$o_1$} } (2);	
		\path (3) edge [->, out=130,in=50,looseness=3] node[above] {\nscale{$o_2$} } (3);	
		\path (4) edge [->, out=130,in=50,looseness=3] node[above] {\nscale{$k$} } (4);
		\graph {
	(import nodes);
			bot5->["\escale{$w_{5}$}"]0->["\escale{$k$}"]1->["\escale{$o_1$}"]2->["\escale{$o_2$}"]3->["\escale{$k$}"]4;
			};
\end{scope}
\begin{scope}[yshift=-7cm,nodes={set=import nodes}]%
		\coordinate (bot6) at (0,0) ;
		\foreach \i in {0,...,2} { \coordinate (\i) at (1.8cm+ \i*1.8cm,0) ;}
		\foreach \i in {bot6} {\fill[green!20, rounded corners] (\i) +(-0.7,-0.35) rectangle +(4,0.7);}
		\node (bot6) at (bot6) {\nscale{$\bot_{6}$}};
		\foreach \i in {0,...,2} { \node (\i) at (\i) {\nscale{$h_{2,\i}$}};}
		\path (0) edge [->, out=130,in=50,looseness=3] node[above] {\nscale{$w_6$} } (0);					
		\path (1) edge [->, out=130,in=50,looseness=3] node[above] {\nscale{$k$} } (1);	
		\path (2) edge [->, out=130,in=50,looseness=3] node[above] {\nscale{$z_1,o_1$} } (2);
		\graph {
	(import nodes);
			bot6->["\escale{$w_{6}$}"]0->["\escale{$k$}"]1->["\escale{$z_1$}"]2;
			};
\end{scope}
\begin{scope}[yshift=-8.4cm,nodes={set=import nodes}]%
		\node (bot7) at (0,0) {\nscale{$\bot_{7}$}};
		\foreach \i in {0} { \coordinate (\i) at (1.8cm+ \i*1.8cm,0) ;}
		\foreach \i in {0} { \node (\i) at (\i) {\nscale{$h_{3,\i}$}};}
		\path (0) edge [->, out=130,in=50,looseness=3] node[above] {\escale{$w_7,o_1,z_2$} } (0);	
		
		\graph {
	(import nodes);
			bot7->["\escale{$w_7$}"]0;
			};
\end{scope}
\path (bot1) edge [->] node[left] {\nscale{$\ominus_1$} } (bot2);
\path (bot2) edge [->] node[left] {\nscale{$\ominus_2$} } (bot3);
\path (bot3) edge [->] node[left] {\nscale{$\ominus_3$} } (bot4);
\path (bot4) edge [->] node[left] {\nscale{$\ominus_4$} } (bot5);
\path (bot5) edge [->] node[left] {\nscale{$\ominus_5$} } (bot6);
\path (bot6) edge [->] node[left] {\nscale{$\ominus_6$} } (bot7);

\path (bot2) edge [->, out=150,in=-150,looseness=3] node[left] {\escale{$\ominus_1$}} (bot2);
\path (bot3) edge [->, out=150,in=-150,looseness=3] node[left] {\escale{$\ominus_2$}} (bot3);
\path (bot4) edge [->, out=150,in=-150,looseness=3] node[left] {\escale{$\ominus_3$}} (bot4);
\path (bot5) edge [->, out=150,in=-150,looseness=3] node[left] {\escale{$\ominus_4$}} (bot5);
\path (bot6) edge [->, out=150,in=-150,looseness=3] node[left] {\escale{$\ominus_5$}} (bot6);
\path (bot7) edge [->, out=150,in=-150,looseness=3] node[left] {\escale{$\ominus_6$}} (bot7);
\end{tikzpicture}
\end{center}
\caption{The TS $A^\tau_I$ where $\tau$ corresponds to Theorem~\ref{the:w2_result}.\ref{the:w2_result_nop_set_res} and $I$ to Example~\ref{ex:hitting_set} with the HS $S=\{X_1,X_3\}$.
The green colored area sketches the $\tau$-region $R=(sup, sig)$ that solves $\alpha$, where, for all $e\in E(A^\tau_I)$, if $e=k$, then $sig(e)=\used$; 
if $e\in \{o_2\}\cup S$, then $sig(e)=\set$; 
if $e\in \{o_1,z_1, \ominus_6\}$, then $sig(e)=\res$; 
otherwise $sig(e)=\nop$.
 }\label{ex:nop_set_res}
\end{figure}

\textbf{Theorem~\ref{the:w2_result}.\ref{the:w2_result_nop_set_res}: The Reduction.}
Let $\tau$ be a type of Theorem~\ref{the:w2_result}.\ref{the:w2_result_nop_set_res}.
According to our general approach, we first define $d=\kappa+4$.
Next we introduce the TS $A^\tau_I$. 
Figure~\ref{ex:nop_set_res} provides an example of $A^\tau_I$, where $I$ corresponds to Example~\ref{ex:hitting_set}.
The TS $A^\tau_I$ has the following gadget $H_1$ that provides the atom $\alpha=(k,h_{1,2})$:
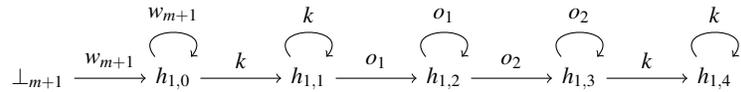
\begin{center}
\begin{tikzpicture}
		\node (b) at (0,0) {\nscale{$\bot_{m+1}$}};
		\foreach \i in {0,...,4} { \coordinate (\i) at (1.8cm+ \i*1.8cm,0) ;}
		\foreach \i in {0,...,4} { \node (\i) at (\i) {\nscale{$h_{1,\i}$}};}
		\path (0) edge [->, out=130,in=50,looseness=3] node[above] {\nscale{$w_{m+1}$} } (0);					
		\path (1) edge [->, out=130,in=50,looseness=3] node[above] {\nscale{$k$} } (1);	
		\path (2) edge [->, out=130,in=50,looseness=3] node[above] {\nscale{$o_1$} } (2);	
		\path (3) edge [->, out=130,in=50,looseness=3] node[above] {\nscale{$o_2$} } (3);	
		\path (4) edge [->, out=130,in=50,looseness=3] node[above] {\nscale{$k$} } (4);
		\graph {
	(import nodes);
			(b)->["\escale{$w_{m+1}$}"]0->["\escale{$k$}"]1->["\escale{$o_1$}"]2->["\escale{$o_2$}"]3->["\escale{$k$}"](4);
			};
\end{tikzpicture}
\end{center}
Moreover, the TS $A^\tau_I$ has the following gadgets $H_2$ and $H_3$:
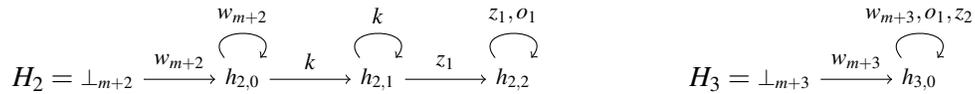
\begin{center}
\begin{tikzpicture}[new set = import nodes]
\begin{scope}[nodes={set=import nodes}]%
		\node (h2) at (-0.85,0) {$H_2=$};
		\node (bot6) at (0,0) {\nscale{$\bot_{m+2}$}};
		\foreach \i in {0,...,2} { \coordinate (\i) at (1.8cm+ \i*1.8cm,0) ;}
		\foreach \i in {0,...,2} { \node (\i) at (\i) {\nscale{$h_{2,\i}$}};}
		\path (0) edge [->, out=130,in=50,looseness=3] node[above] {\nscale{$w_{m+2}$} } (0);					
		\path (1) edge [->, out=130,in=50,looseness=3] node[above] {\nscale{$k$} } (1);	
		\path (2) edge [->, out=130,in=50,looseness=3] node[above] {\nscale{$z_1,o_1$} } (2);
		\graph {
	(import nodes);
			bot6->["\escale{$w_{m+2}$}"]0->["\escale{$k$}"]1->["\escale{$z_1$}"]2;
			};
\end{scope}
\begin{scope}[xshift=9cm,nodes={set=import nodes}]%
		\node (h3) at (-0.85,0) {$H_3=$};
		\node (bot7) at (0,0) {\nscale{$\bot_{m+3}$}};
		\foreach \i in {0} { \coordinate (\i) at (1.8cm+ \i*1.8cm,0) ;}
		\foreach \i in {0} { \node (\i) at (\i) {\nscale{$h_{3,\i}$}};}
		\path (0) edge [->, out=130,in=50,looseness=3] node[above] {\escale{$w_{m+3},o_1,z_2$} } (0);	
		
		\graph { (import nodes); 	bot7->["\escale{$w_{m+3}$}"]0; 	};
\end{scope}

\end{tikzpicture}
\end{center}
For all $i\in \{1,\dots, m\}$, TS $A^\tau_I$ has the following gadget $T_i$ that applies $w_i,k,z_1,z_2$ and the elements of $M_i=\{X_{i_1},\dots, X_{i_{m_i}}\}$ as events: 
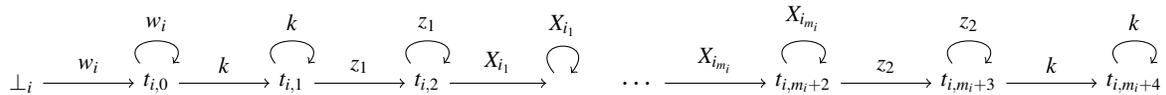
\begin{center}
\begin{tikzpicture}[new set = import nodes]
\begin{scope}[nodes={set=import nodes}]%
		\coordinate (0) at (0,0);
		\coordinate (1) at (1.8,0);
		\coordinate (2) at (3.6,0);
		\coordinate (3) at (5.4,0);
		\coordinate (31) at (7.2,0);
		\coordinate (dots) at (8.2,0);
		\coordinate (4) at (10.4,0);
		\coordinate (5) at (12.6,0);
		\coordinate (6) at (14.8,0);

		\node (0) at (0) {\nscale{$\bot_i$}};
		\node (1) at (1) {\nscale{$t_{i,0}$}};	
		\node (2) at (2) {\nscale{$t_{i,1}$}};	
		\node (3) at (3) {\nscale{$t_{i,2}$}};
		\node (31) at (31) {$ \; \;$};		
		\node (dots) at (dots) {$\dots$};
		\node (4) at (4) {\nscale{$t_{i,m_i+2}$}};
		\node (5) at (5) {\nscale{$t_{i,m_i+3}$}};
		\node (6) at (6) {\nscale{$t_{i,m_i+4}$}};	
		
		\path (1) edge [->, out=130,in=50,looseness=3] node[above] {\escale{$w_i$} } (1);
		\path (2) edge [->, out=130,in=50,looseness=3] node[above] {\escale{$k$} } (2);
		\path (3) edge [->, out=130,in=50,looseness=3] node[above] {\escale{$z_1$} } (3);
		\path (31) edge [->, out=130,in=50,looseness=7] node[above] {\escale{$X_{i_1}$} } (31);
		\path (4) edge [->, out=130,in=50,looseness=3] node[above] {\escale{$X_{i_{m_i}}$} } (4);
		\path (5) edge [->, out=130,in=50,looseness=3] node[above] {\escale{$z_2$} } (5);	
		\path (6) edge [->, out=130,in=50,looseness=3] node[above] {\escale{$k$} } (6);				
		\graph {
	(import nodes);
			0 ->["\escale{$w_i$}"]1->["\escale{$k$}"]2->["\escale{$z_1$}"]3->["\escale{$X_{i_1}$}"]31;
			dots->["\escale{$X_{i_{m_i}}$}"]4->["\escale{$z_2$}"]5->["\escale{$k$}"]6;
			};
\end{scope}
\end{tikzpicture}
\end{center}
Finally, the TS $A^\tau_I$ uses the events $\ominus_1,\dots, \ominus_{m+2}$ and applies for all $i\in \{1,\dots, m\}$ the edges $\bot_i\edge{\ominus_i}\bot_{i+1}$ and $\bot_{i+1}\edge{\ominus_i}\bot_{i+1}$ to join the gadgets $T_1,\dots, T_m$ and $H_1$, $H_2$, $H_3$.

\textbf{Theorem~\ref{the:w2_result}.\ref{the:w2_result_nop_set_res}: The $\tau$-solvability of $\alpha$ Implies a Hitting Set.} 
Let $R=(sup, sig)$ be a $\tau$-region that solves $\alpha$, that is, either $sig(k)=\used$ and $sup(h_{1,2})=0$ or $sig(k)=\free$ and $sup(h_{1,2})=1$.
In the following, we assume that $sig(k)=\used$ and $sup(h_{1,2})=0$.
The arguments for the case $sig(k)=\free$ and $sup(h_{1,2})=1$ are symmetrical.
Notice that if $s\edge{e}s'\in A^\tau_I$, then $s'\edge{e}s'\in A^\tau_I$.
Thus, for all $e\in E(A^\tau_I)$ holds $sig(e)\not=\swap$.

Since $sig(k)=\used$, if $s\edge{k}s'$, then $sup(s)=sup(s')=1$.
In particular, we have $sup(t_{i,m_i+3})=1$ for all $i\in \{1,\dots, m\}$.
Moreover, by $sup(h_{1,1})=1$ and $sup(h_{1,2})=0$, we have $sig(o_1)=\res$ and $sig(o_2)=\set$.
This implies $sup(h_{2,2})=sup(h_{3,0})=0$.
By $sup(h_{2,1})=1$ and $sup(h_{2,2})=0$, we get $sig(z_1)=\res$;
by $sup(h_{3,0})=0$ and $sup(t_{1,m_i+3})=1$, we get $sig(z_2)=\nop$.
Thus, by $sig(z_1)=\res$ and $sig(z_2)=\nop$, we get $sup(t_{i,2})=0$ and $sup(t_{i,m_i+2})=1$ for all $i\in \{1,\dots, m\}$.
Consequently, for all $i\in \{1,\dots, m\}$, there is $X\in M_i$ such that $sig(X)=\set$.
Since $sig(e)\not=\nop$ for all $e\in \{k,o_1,o_2,z_1\}$ and $R$ is $d$-restricted, it holds $\vert \{X\in \mathfrak{U}\mid sig(X)\not=\nop\}\vert \leq \kappa $.
This implies that $S= \{X\in \mathfrak{U}\mid sig(X)\not=\nop\}$ is a sought-for hitting set of $I$.

In return, if $(\mathfrak{U}, M,\kappa)$ has a $\kappa$-HS, then $A_\tau^I$ is $\tau$-solvable, which is the statement of the following lemma.
Due to space restrictions, we omit the proof which can be found in \cite{DBLP:journals/corr/abs-2007-12372}.
\begin{lemma}\label{lem:w2_result_nop_set_res}
Let $\tau$ be a type of nets in correspondence of Theorem~\ref{the:w2_result}.\ref{the:w2_result_nop_set_res}.
If $(\mathfrak{U}, M,\kappa)$ has a $\kappa$-HS, then there is a $d$-restricted admissible set of $A_\tau^I$.
\end{lemma}

\subsection{Proof of Theorem~\ref{the:w2_result}.\ref{the:w2_result_nop_set_swap}}\label{app:w2_result_nop_set_swap}%

\textbf{Theorem~\ref{the:w2_result}.\ref{the:w2_result_nop_set_swap}: The Reduction.}
We restrict ourselves to the case where $\tau=\{\nop,\set,\swap\}\cup\omega$ or $\tau=\{\nop,\out,\set,\swap\}\cup\omega$ and $\emptyset\not=\omega\subseteq\{\free,\used\}$.
The hardness for the other types follows by symmetry.
First, we define $d=\kappa+4$.
Next, we introduce the TS $A^\tau_I$.
Figure~\ref{fig:example_nop_set_swap_used} provides a full example of $A^\tau_I$ where $I$ corresponds to Example~\ref{ex:hitting_set}.

The TS $A^\tau_I$ has the following gadgets $H_0$ and $H_1$ that provide the atom $\alpha=(k,h_{0,3})$:
\begin{center}
\begin{tikzpicture}[new set = import nodes]
\begin{scope}[nodes={set=import nodes}]%

		\coordinate (bot0) at (0,0);
		\foreach \i in {1,...,5} { \coordinate (\i) at (\i*1.8cm,0) ;}
		\node (H0) at (-0.9,0) {$H_0=$};
		\node (bot0) at (0,0) {\nscale{$\bot_{m+1}$}};
		\foreach \i in {1,...,5} { \node (\i) at (\i) {\nscale{$h_{0,\i}$}};}
		\graph {
	(import nodes);
			bot0 <->["\escale{$w_{m+1}$}"]1<->["\escale{$k$}"]2<->["\escale{$o_1$}"]3<->["\escale{$o_2$}"]4<->["\escale{$k$}"]5;
		};
\end{scope}
\begin{scope}[yshift=-1.2cm,nodes={set=import nodes}]%
		
		\coordinate (bot1) at (0,0);
		\foreach \i in {1,...,6} { \coordinate (\i) at (\i*1.8cm,0) ;}
		\node (H1) at (-0.9,0) {$H_1=$};
		\node (bot1) at (0,0) {\nscale{$\bot_{m+2}$}};
		\foreach \i in {1,...,6} { \node (\i) at (\i) {\nscale{$h_{1,\i}$}};}
\graph {
	(import nodes);
			bot1 <->["\escale{$w_{m+2}$}"]1<->["\escale{$k$}"]2<->["\escale{$z_1$}"]3<->["\escale{$o_1$}"]4<->["\escale{$z_2$}"]5<->["\escale{$k$}"]6;
		};
\end{scope}
\end{tikzpicture}
\end{center}

Moreover, for every $i\in \{1,\dots, m\}$, the TS $A^\tau_I$ has the following gadget $T_i$ that has the elements of $M_i=\{X_{i_1},\dots, X_{i_{m_i}}\}$ as events:
\begin{center}
\begin{tikzpicture}[new set = import nodes]
\begin{scope}[nodes={set=import nodes}]%

		\coordinate (boti) at (0,0);
		\foreach \i in {0,...,6} { \coordinate (\i) at (\i*2cm+2cm,0) ;}
		\node (boti) at (boti) {\nscale{$\bot_i$}};
		\foreach \i in {0,...,6} { \node (\i) at (\i) {\nscale{$t_{i,\i}$}};}
		\node (dots) at (14,-0.6) {\nscale{$\vdots$}};
		\node (t0) at (14,-1.2) {\nscale{$t_{i,4m_i-2}$}};
		\node (t1) at (11.5,-1.2) {\nscale{$t_{i,4m_i-1}$}};
		\node (t2) at (9,-1.2) {\nscale{$t_{i,4m_i}$}};
		\node (t3) at (6.5,-1.2) {\nscale{$t_{i,4m_i+1}$}};
		\node (t4) at (4,-1.2) {\nscale{$t_{i,4m_i+2}$}};
		\node (t5) at (1.5,-1.2) {\nscale{$t_{i,4m_i+3}$}};
		\node (t6) at (-0.75,-1.2) {\nscale{$t_{i,4m_i+4}$}};
		\graph {
	(import nodes);
			boti <->["\escale{$w_i$}"]0<->["\escale{$k$}"]1<->["\escale{$z_1$}"]2<->["\escale{$a_{i,1}$}"]3->["\escale{$X_{i_1}$}"]4<->["\escale{$X_{i_1}$}"]5<->["\escale{$a_{i,1}$}"]6;
			t0<->[swap, "\escale{$a_{i,m_i}$}"]t1->[swap,"\escale{$X_{i_{m_i}}$}"]t2<->[swap,"\escale{$X_{i_{m_i}}$}"]t3<->[swap,"\escale{$a_{i,m_i}$}"]t4<->[swap,"\escale{$z_2$}"]t5<->[swap,"\escale{$k$}"]t6;
		};
\end{scope}
\end{tikzpicture}
\end{center}
Notice that, for all $\ell\in \{1,\dots, m_i\}$, the event $a_{i,\ell}$ that encompasses the event $X_{i_\ell}$ of $M_i$ is bounded to the occurrence of $X_{i_\ell}$ in $T_i$.
In particular, if two distinct sets $M_i$ and $M_j$ share an event $X\in \mathfrak{U}$, that is, there are indices $\ell\in \{1,\dots, m_i\}$ and $n\in \{1,\dots, m_j\}$ such that $X=X_{i_\ell}=X_{j_n}$, then $a_{i,\ell}$ embraces $X$ in $T_i$ and $a_{j,n}$ embraces $X$ in $T_j$ but $a_{i,\ell}$ and $a_{j,n}$ are distinct.
Finally, to obtain $A^\tau_I$, we use fresh events $\ominus_1,\dots, \ominus_{m+1}$ and connect $T_1,\dots, T_m, H_0$ and $H_1$ by $\bot_1\fbedge{\ominus_1}\dots\fbedge{\ominus_{m+1}}\bot_{m+2}$. 
The initial state of $A^\tau_I$ is $\bot_1$.
Notice that for every region $R$ of $A^\tau_I$, holds that $s\fbedge{e}s'\in A^\tau_I$ and $sup(s)\not=sup(s')$ implies $sig(e)=\swap$.
Moreover, if $s\edge{e}s'\in A^\tau_I$, then, by construction, $s'\edge{e}$.
By the definition of \out, this implies $sig(e)\not=\out$ for all $e\in E(A^\tau_I)$.

\begin{figure}[t!]
\begin{center}
\begin{tikzpicture}[new set = import nodes]
\begin{scope}[nodes={set=import nodes}]%

		\coordinate (bot1) at (0,0);
		\foreach \i in {0,...,7} { \coordinate (\i) at (\i*1.5cm+1.5cm,0);}
		\foreach \i in {8,...,12} {\pgfmathparse{int(\i-8)} \coordinate (\i) at (12cm-\pgfmathresult*1.5cm,-1.2cm);}
		
		\foreach \i in {bot1} {\fill[green!20, rounded corners] (\i) +(-0.4,-0.25) rectangle +(3.5,0.3);}
		\foreach \i in {4} {\fill[green!20, rounded corners] (\i) +(-0.4,-0.25) rectangle +(4.5,0.3);}
		\foreach \i in {7} {\fill[green!20, rounded corners] (\i) +(-0.2,-1.2) rectangle +(0.5,0.3);}
		\foreach \i in {12} {\fill[green!20, rounded corners] (\i) +(-0.3,-0.4) rectangle +(6.5,0.3);}
		\foreach \i in {bot1} {\fill[green!20, rounded corners] (\i) +(-0.4,-11.5) rectangle +(0.3,0.3);}
		\node (bot1) at (bot1) {\nscale{$\bot_1$}};
		\foreach \i in {0,...,12} { \node (\i) at (\i) {\nscale{$t_{1,\i}$}};}
\graph {
	(import nodes);
			bot1 <->["\escale{$w_1$}"]0<->["\escale{$k$}"]1<->["\escale{$z_1$}"]2<->["\escale{$a_{1,1}$}"]3->["\escale{$X_1$}"]4<->["\escale{$X_1$}"]5<->["\escale{$a_{1,1}$}"]6<->["\escale{$a_{1,2}$}"]7->["\escale{$X_2$}"]8<->["\escale{$X_2$}"]9<->["\escale{$a_{1,2}$}"]10<->["\escale{$z_2$}"]11<->["\escale{$k$}"]12;  
			};
\end{scope}
\begin{scope}[yshift=-2.5cm,nodes={set=import nodes}]%

		\coordinate (bot2) at (0,0);
		\foreach \i in {0,...,7} { \coordinate (\i) at (\i*1.5cm+1.5cm,0);}
		\foreach \i in {8,...,12} {\pgfmathparse{int(\i-8)} \coordinate (\i) at (12cm-\pgfmathresult*1.5cm,-1.2cm);}
		\foreach \i in {bot2} {\fill[green!20, rounded corners] (\i) +(-0.4,-0.25) rectangle +(3.5,0.3);}
		\foreach \i in {12} {\fill[green!20, rounded corners] (\i) +(-0.3,-0.4) rectangle +(6.5,0.3);}
		\node (bot2) at (bot2) {\nscale{$\bot_2$}};
		\foreach \i in {0,...,12} { \node (\i) at (\i) {\nscale{$t_{2,\i}$}};}
\graph {
	(import nodes);
			bot2 <->["\escale{$w_2$}"]0<->["\escale{$k$}"]1<->["\escale{$z_1$}"]2<->["\escale{$a_{2,1}$}"]3->["\escale{$X_2$}"]4<->["\escale{$X_2$}"]5<->["\escale{$a_{2,1}$}"]6<->["\escale{$a_{2,2}$}"]7->["\escale{$X_3$}"]8<->["\escale{$X_3$}"]9<->["\escale{$a_{2,2}$}"]10<->["\escale{$z_2$}"]11<->["\escale{$k$}"]12;  
			};
\end{scope}
\begin{scope}[yshift=-5cm,nodes={set=import nodes}]%

		\coordinate (bot3) at (0,0);
		\foreach \i in {0,...,7} { \coordinate (\i) at (\i*1.5cm+1.5cm,0);}
		\foreach \i in {8,...,12} {\pgfmathparse{int(\i-8)} \coordinate (\i) at (12cm-\pgfmathresult*1.5cm,-1.2cm);}
		\foreach \i in {bot3} {\fill[green!20, rounded corners] (\i) +(-0.4,-0.25) rectangle +(3.5,0.3);}
		\foreach \i in {4} {\fill[green!20, rounded corners] (\i) +(-0.4,-0.25) rectangle +(4.5,0.3);}
		\foreach \i in {7} {\fill[green!20, rounded corners] (\i) +(-0.2,-1.2) rectangle +(0.5,0.3);}
		\foreach \i in {12} {\fill[green!20, rounded corners] (\i) +(-0.3,-0.4) rectangle +(6.5,0.3);}
		\node (bot3) at (bot3) {\nscale{$\bot_3$}};
		\foreach \i in {0,...,12} { \node (\i) at (\i) {\nscale{$t_{3,\i}$}};}
\graph {
	(import nodes);
			bot3 <->["\escale{$w_3$}"]0<->["\escale{$k$}"]1<->["\escale{$z_1$}"]2<->["\escale{$a_{3,1}$}"]3->["\escale{$X_1$}"]4<->["\escale{$X_1$}"]5<->["\escale{$a_{3,1}$}"]6<->["\escale{$a_{3,2}$}"]7->["\escale{$X_4$}"]8<->["\escale{$X_4$}"]9<->["\escale{$a_{3,2}$}"]10<->["\escale{$z_2$}"]11<->["\escale{$k$}"]12;  
			};
\end{scope}
\begin{scope}[yshift=-7.5cm,nodes={set=import nodes}]%

		\coordinate (bot4) at (0,0);
		\foreach \i in {0,...,8} { \coordinate (\i) at (\i*1.5cm+1.5cm,0);}
		\foreach \i in {9,...,16} {\pgfmathparse{int(\i-9)} \coordinate (\i) at (13.5cm-\pgfmathresult*1.5cm,-1.2cm);}
		
		\foreach \i in {bot4} {\fill[green!20, rounded corners] (\i) +(-0.4,-0.25) rectangle +(3.5,0.3);}
		\foreach \i in {4} {\fill[green!20, rounded corners] (\i) +(-0.4,-0.25) rectangle +(6,0.3);}
		\foreach \i in {8} {\fill[green!20, rounded corners] (\i) +(-0.2,-1.2) rectangle +(0.5,0.3);}
		\foreach \i in {16} {\fill[green!20, rounded corners] (\i) +(-0.3,-0.4) rectangle +(11,0.3);}
		\node (bot4) at (bot4) {\nscale{$\bot_4$}};
		\foreach \i in {0,...,16} { \node (\i) at (\i) {\nscale{$t_{4,\i}$}};}
\graph {
	(import nodes);
			bot4 <->["\escale{$w_4$}"]0<->["\escale{$k$}"]1<->["\escale{$z_1$}"]2<->["\escale{$a_{4,1}$}"]3->["\escale{$X_1$}"]4<->["\escale{$X_1$}"]5<->["\escale{$a_{4,1}$}"]6<->["\escale{$a_{4,2}$}"]7->["\escale{$X_3$}"]8<->["\escale{$X_3$}"]9<->["\escale{$a_{4,2}$}"]10<->["\escale{$a_{4,3}$}"]11->["\escale{$X_4$}"]12<->["\escale{$X_4$}"]13<->["\escale{$a_{4,3}$}"]14    <->["\escale{$z_2$}"]15<->["\escale{$k$}"]16;  
			};
\end{scope}
\begin{scope}[yshift=-10cm,nodes={set=import nodes}]%

		\coordinate (bot5) at (0,0);
		\foreach \i in {0,...,4} { \coordinate (\i) at (\i*1.5cm+1.5cm,0);}
		\foreach \i in {bot5} {\fill[green!20, rounded corners] (\i) +(-0.4,-0.25) rectangle +(3.5,0.3);}
		\foreach \i in {3} {\fill[green!20, rounded corners] (\i) +(-0.4,-0.25) rectangle +(1.75,0.3);}
		\foreach \i in {0,...,4} { \node (\i) at (\i) {\nscale{$h_{0,\i}$}};}
		\node (bot5) at (0,0) {\nscale{$\bot_5$}};
		\graph {
	(import nodes);
			bot5 <->["\escale{$w_5$}"]0<->["\escale{$k$}"]1<->["\escale{$o_1$}"]2<->["\escale{$o_2$}"]3<->["\escale{$k$}"]4;
		};
\end{scope}
\begin{scope}[yshift=-11.5cm,nodes={set=import nodes}]%
		
		\coordinate (bot6) at (0,0);
		\foreach \i in {0,...,5} { \coordinate (\i) at (\i*1.5cm+1.5cm,0);}
		\foreach \i in {bot6} {\fill[green!20, rounded corners] (\i) +(-0.4,-0.25) rectangle +(3.5,0.3);}
		\foreach \i in {3} {\fill[green!20, rounded corners] (\i) +(-0.4,-0.25) rectangle +(3.25,0.3);}
		\foreach \i in {0,...,5} { \node (\i) at (\i) {\nscale{$h_{1,\i}$}};}
		\node (bot6) at (0,0) {\nscale{$\bot_6$}};
		\graph {
	(import nodes);
			bot6 <->["\escale{$w_6$}"]0<->["\escale{$k$}"]1<->["\escale{$z_1$}"]2<->["\escale{$o_1$}"]3<->["\escale{$z_2$}"]4<->["\escale{$k$}"]5;
		};
\end{scope}

\path (bot1) edge [<->] node[left] {\nscale{$\ominus_1$} } (bot2);
\path (bot2) edge [<->] node[left] {\nscale{$\ominus_2$} } (bot3);
\path (bot3) edge [<->] node[left] {\nscale{$\ominus_3$} } (bot4);
\path (bot4) edge [<->] node[left] {\nscale{$\ominus_4$} } (bot5);
\path (bot5) edge [<->] node[left] {\nscale{$\ominus_5$} } (bot6);

\end{tikzpicture}
\end{center}
\caption{
A full example of $A^\tau_I$, where $\tau$ belongs to the types of Theorem~\ref{the:w2_result}.\ref{the:w2_result_nop_set_swap} and $I$ originates from Example~\ref{ex:hitting_set}.
Green colored area: A sketch of the $\{\nop,\set,\swap,\used\}$-region $R^k=(sup, sig)$, based on the HS $S=\{X_1,X_3\}$, that satisfies $sig(k)=\used$ and $sup(h_{0,2})=0$ and solves $\alpha$.}\label{fig:example_nop_set_swap_used}
\end{figure}

\textbf{Theorem~\ref{the:w2_result}.\ref{the:w2_result_nop_set_swap}: The $\tau$-Solvability of $\alpha$ Implies a Hitting Set.}
Let $R=(sup, sig)$ be a $\tau$-region that solves $\alpha$.
Since $R$ solves $\alpha$, we have either $sig(k)=\used$ and $sup(h_{0,3})=0$ or $sig(k)=\free$ and $sup(h_{0,3})=1$.
In the following, we consider the former case, the arguments for the latter are symmetrical.
Please note Figure~\ref{fig:example_nop_set_swap_used} during the following considerations.
By $sig(k)=\used$, we have that $sup(s)=sup(s')=1$ for all $s\edge{k}s'\in A^\tau_I$.
In particular, we have $sup(h_{0,2})=sup(h_{0,4})=1$ which, by $sup(h_{0,3}) = 0$, implies $sig(o_1)=sig(o_2)=\swap$.
Moreover, we have $sup(h_{1,2})=sup(h_{1,5})=1$.
Consequently, the number of state changes on the image $P^R$ of the path $P=h_{1,2}\edge{z_1}\dots\edge{z_2}h_{1,5}$ is even.
Since $sig(o_1)=\swap$, this implies that there is exactly one event $e\in \{z_1,z_2\}$ such that $sig(e)=\swap$.
We consider the case $sig(z_1)=\swap$.
The arguments for the case $sig(z_2)=\swap$ are similar.
The region $R$ is $d$-restricted, and $k,o_1,o_2, z_1$ have signatures different from \nop.
There are at most $\kappa$ events left whose signatures are not \nop. 

Let $i\in \{1,\dots, m\}$ be arbitrary but fixed.
By $sig(k)=\used$, we have $sup(t_{i,1})=sup(t_{i,4m_i+3})=1$.
By $sig(z_1)=\swap$ and $sig(z_2)\not=\swap$, this implies $sup(t_{i,2})=0$ and $sup(t_{i,m_i+2})=1$.
Hence the image $P^R$ of the path $P=$
\begin{center}
\begin{tikzpicture}[new set = import nodes]
\begin{scope}[nodes={set=import nodes}]%
		\foreach \i in {0,...,4} { \coordinate (\i) at (\i*1.3cm,0) ;}
		\foreach \i in {6,...,10} { \pgfmathparse{int(\i-6)}  \coordinate (\i) at (\pgfmathresult*1.9cm+7cm,0) ;}
		\foreach \i in {0,...,4} { \pgfmathparse{int(\i+2)} \node (\i) at (\i) {\nscale{$t_{i,\pgfmathresult}$}};}
		\node (dots) at (6,0) {$\dots$};
		\node (4m_2) at (6) {\nscale{$t_{i,4m_i-2}$}};
		\node (4m_1) at (7) {\nscale{$t_{i,4m_i-1}$}};
		\node (4m) at (8) {\nscale{$t_{i,4m_i}$}};
		\node (4m+1) at (9) {\nscale{$t_{i,4m_i+1}$}};
		\node (4m+2) at (10) {\nscale{$t_{i,4m_i+2}$}};
		\graph {
	(import nodes);
			0 <->["\escale{$a_{i,1}$}"]1->["\escale{$X_{i_1}$}"]2<->["\escale{$X_{i_1}$}"]3<->["\escale{$a_{i,1}$}"]4;
			4m_2 <->["\escale{$a_{i,m_i}$}"]4m_1->["\escale{$X_{i_{m_i}}$}"]4m<->["\escale{$X_{i_{m_i}}$}"]4m+1<->["\escale{$a_{i,m_i}$}"]4m+2;
			};
\end{scope}
\end{tikzpicture}
\end{center}
is a path from $0$ to $1$ in $\tau$.
Thus, there is an event $e\in \{X_{i_1},\dots, X_{i_{m_i}}\}\cup\{a_{i,1},\dots, a_{i,m_i}\}$ whose signature causes the state change from $0$ to $1$.
This implies $sig(e)\not=\nop$.
Assume, for a contradiction, that $sig(e)=\nop$ for all $e\in \{X_{i_1},\dots, X_{i_{m_i}}\}$.
Let $\ell\in \{1,\dots, m_i\}$ be arbitrary but fixed.
By $sig(X_{\ell}) = \nop$, we get $sup(t_{i,4\ell-1})=sup(t_{i,4\ell})=sup(t_{i,4\ell+1})$. 
Recall that $sup(s)\not=sup(s')$ implies $sig(e)=\swap$ for all $s\fbedge{e}s'\in A^\tau_I$.
Thus, if $sig(a_{i,\ell})\not=\swap$, then $sup(t_{i,4\ell-2})=sup(t_{i,4\ell-1})=sup(t_{i,4\ell})=sup(t_{i,4\ell+1})=sup(t_{i,4\ell+2})$.
Otherwise, if $sig(a_{i,\ell})=\swap$, then $sup(t_{i,4\ell-2})\not=sup(t_{i,4\ell-1})=sup(t_{i,4\ell})=sup(t_{i,4\ell+1})\not=sup(t_{i,4\ell+2})$.
Consequently, both cases imply $sup(t_{i,4\ell-2})=sup(t_{i,4\ell+2})$.
Since $\ell$ was arbitrary, this implies $sup(t_{i,2})=sup(t_{i,4m_i+2})$, a contradiction.
Hence, there is an event $e\in  \{X_{i_1},\dots, X_{i_{m_i}}\}$ such that $sig(e)\not=\nop$.
Since $i$ was arbitrary, this is simultaneously true for all $T_1,\dots, T_m$.
Moreover, since $R$ respects the parameter, the cardinality of $S=\{X\in \mathfrak{U}\mid sig(X)\not=\nop\}$ is at most $\kappa$. 
Thus, $S$ is a fitting hitting set of $I$.

The next lemma completes the proof of Theorem~\ref{the:w2_result}.\ref{the:w2_result_nop_set_swap} and states that a sought HS of $I$ implies  a $d$-restricted admissible set of $A^\tau_I$.
Due to space restrictions, its proof can be found in~\cite{DBLP:journals/corr/abs-2007-12372}.
\begin{lemma}\label{lem:hs_implies_solvability_nop_set_swap}
Let $\tau$ be a type of net corresponding to Theorem~\ref{the:w2_result}.\ref{the:w2_result_nop_set_swap}.
If $I=(\mathfrak{U}, M, \kappa)$ has a fitting HS, then $A^\tau_I$ has a $d$-restricted admissible set.
\end{lemma}
%

\subsection{The Proof of Theorem~\ref{the:w2_result}.\ref{the:w2_result_nop_inp_res_swap}}%

\textbf{Theorem~\ref{the:w2_result}.\ref{the:w2_result_nop_inp_res_swap}: The Reduction}
In the following, we argue for $\tau=\{\nop,\inp,\res,\swap\}$.
The hardness for $\tau=\{\nop,\out,\set,\swap\}$ then follows by symmetry.
For a start, we define $d=\kappa+4$.
The TS $A^\tau_I$ has the following gadgets $H_0,\dots, H_4$ that provide the atom $\alpha=(k,h_{0,2})$: 
\begin{center}
\begin{tikzpicture}[new set = import nodes]
\begin{scope}[nodes={set=import nodes}]
		
		\node (H) at (-2.75,0) {$H_0=$};
		\node (bot) at (-1.8,0) {$\bot_{m+1}$};
		\foreach \i in {0,...,4} { \coordinate (\i) at (\i*1.8cm,0) ;}
		\foreach \i in {0,...,4} { \node (\i) at (\i) {\nscale{$h_{0,\i}$}};}
\graph {
	(import nodes);
			bot->["\escale{$w_{m+1}$}"] 0 ->["\escale{$k$}"]1->["\escale{$o_1$}"]2->["\escale{$o_2$}"]3->["\escale{$k$}"]4;  
			};
\end{scope}
\begin{scope}[yshift=-1cm, nodes={set=import nodes}]
		
		\node (H) at (-2.75,0) {$H_1=$};
		\node (bot) at (-1.8,0) {$\bot_{m+2}$};
		\foreach \i in {0,...,4} { \coordinate (\i) at (\i*1.8cm,0) ;}
		\foreach \i in {0,...,4} { \node (\i) at (\i) {\nscale{$h_{1,\i}$}};}
\graph {
	(import nodes);
			bot->["\escale{$w_{m+2}$}"] 0 ->["\escale{$k$}"]1->["\escale{$z_1$}"]2->["\escale{$o_2$}"]3->["\escale{$k$}"]4;  
			};
\end{scope}
\begin{scope}[yshift=-2cm,nodes={set=import nodes}]
		\node (H) at (-2.75,0) {$H_2=$};
		\node (bot) at (-1.8,0) {$\bot_{m+3}$};
		\foreach \i in {0,...,4} { \coordinate (\i) at (\i*1.8cm,0) ;}		
		\foreach \i in {0,...,4} { \node (\i) at (\i) {\nscale{$h_{2,\i}$}};}
\graph {
	(import nodes);
			bot->["\escale{$w_{m+3}$}"] 0 ->["\escale{$k$}"]1->["\escale{$z_2$}"]2->["\escale{$o_2$}"]3->["\escale{$k$}"]4;  
			};
\end{scope}
\begin{scope}[yshift=-3cm,nodes={set=import nodes}]
		\node (H) at (-2.75,0) {$H_3=$};
		\node (bot) at (-1.8,0) {$\bot_{m+4}$};
		\foreach \i in {0,...,5} { \coordinate (\i) at (\i*1.8cm,0) ;}
		\foreach \i in {0,...,5} { \node (\i) at (\i) {\nscale{$h_{3,\i}$}};}
\graph {
	(import nodes);
			bot->["\escale{$w_{m+4}$}"] 0 ->["\escale{$k$}"]1->["\escale{$z_1$}"]2->["\escale{$z_3$}"]3->["\escale{$z_2$}"]4->["\escale{$k$}"]5;  
			};
\end{scope}
\begin{scope}[yshift=-4cm,nodes={set=import nodes}]
		\node (H) at (-2.75,0) {$H_4=$};
		\node (bot) at (-1.8,0) {$\bot_{m+5}$};
		\foreach \i in {0,...,5} { \coordinate (\i) at (\i*1.8cm,0) ;}
		\foreach \i in {0,...,5} { \node (\i) at (\i) {\nscale{$h_{4,\i}$}};}
\graph {
	(import nodes);
			bot->["\escale{$w_{m+5}$}"] 0 ->["\escale{$k$}"]1->["\escale{$z_1$}"]2->["\escale{$z_4$}"]3->["\escale{$z_2$}"]4->["\escale{$k$}"]5;  
			};
\end{scope}

\end{tikzpicture}
\end{center}
Moreover, for every $i\in \{1,\dots, m\}$, the TS $A^\tau_I$ has the following gadget $T_i$ that uses the elements of $M_i=\{X_{i_1},\dots, X_{i_{m_i}}\}$ as events: 
\begin{center}
\begin{tikzpicture}[new set = import nodes]
\begin{scope}[nodes={set=import nodes}]%
		\coordinate (0) at (0,0);
		\coordinate (1) at (2,0);
		\coordinate (2) at (4,0);
		\coordinate (dots) at (6,0);
		\coordinate (3) at (8.25,0);
		\coordinate (4) at (10.5,0);
		\coordinate (5) at (13,0);
		\foreach \i in {0,...,2} { \node (\i) at (\i) {\nscale{$t_{i,\i}$}};}
		\node (dots) at (dots) {$\dots$};
		\node (3) at (3) {\nscale{$t_{i,m_i+2}$}};
		\node (4) at (4) {\nscale{$t_{i,m_i+3}$}};
		\node (5) at (5) {\nscale{$t_{i,m_i+4}$}};
\graph {
	(import nodes);
			0 ->["\escale{$k$}"]1->["\escale{$z_3$}"]2->["\escale{$X_{i_1}$}"]dots->["\escale{$X_{i_{m_i}}$}"]3->["\escale{$z_4$}"]4->["\escale{$k$}"]5;  
			};
\end{scope}
\end{tikzpicture}
\end{center}
\begin{figure}
\begin{minipage}{\textwidth}
\begin{center}
\begin{tikzpicture}[new set = import nodes]
\begin{scope}[nodes={set=import nodes}]
		\coordinate (bot1) at (0,0);
		\node (bot1) at (bot1) {\nscale{$\bot_1$}};
		\foreach \i in {0,...,6} { \coordinate (\i) at (\i*1.4cm+2cm,0) ;}
		\foreach \i in {2} {\fill[red!20, rounded corners] (\i) +(-0.4,-0.25) rectangle +(0.4,0.4);}
		\foreach \i in {3} {\fill[green!20, rounded corners] (\i) +(-0.4,-0.25) rectangle +(4.6,0.4);}
		\foreach \i in {0,...,6} { \node (\i) at (\i) {\nscale{$t_{1,\i}$}};}
\graph {
	(import nodes);
			bot1->[snake]0 ->["\escale{$k$}"]1->["\escale{$z_3$}"]2->["\escale{$X_{1}$}"]3->["\escale{$X_{2}$}"]4->["\escale{$z_4$}"]5->["\escale{$k$}"]6;  
			};
\end{scope}
\begin{scope}[yshift=-1.2cm,nodes={set=import nodes}]
		\coordinate (bot2) at (0,0);
		\node (bot2) at (bot2) {\nscale{$\bot_2$}};
		\foreach \i in {0,...,6} { \coordinate (\i) at (\i*1.4cm+2cm,0) ;}
		\foreach \i in {2} {\fill[red!20, rounded corners] (\i) +(-0.4,-0.25) rectangle +(6,0.4);}
		\foreach \i in {0,...,6} { \node (\i) at (\i) {\nscale{$t_{2,\i}$}};}
\graph {
	(import nodes);
			bot2->[snake]0 ->["\escale{$k$}"]1->["\escale{$z_3$}"]2->["\escale{$X_{2}$}"]3->["\escale{$X_{3}$}"]4->["\escale{$z_4$}"]5->["\escale{$k$}"]6;  
			};
\end{scope}
\begin{scope}[yshift=-2.4cm,nodes={set=import nodes}]
		\coordinate (bot3) at (0,0);
		\node (bot3) at (bot3) {\nscale{$\bot_3$}};
		\foreach \i in {0,...,6} { \coordinate (\i) at (\i*1.4cm+2cm,0) ;}
		\foreach \i in {3} {\fill[green!20, rounded corners] (\i) +(-0.4,-0.25) rectangle +(0.4,0.3);}
		\foreach \i in {2} {\fill[red!20, rounded corners] (\i) +(-0.4,-0.25) rectangle +(0.4,0.4);}
		\foreach \i in {0,...,6} { \node (\i) at (\i) {\nscale{$t_{3,\i}$}};}
\graph {
	(import nodes);
			bot3->[snake]0 ->["\escale{$k$}"]1->["\escale{$z_3$}"]2->["\escale{$X_{1}$}"]3->["\escale{$X_{4}$}"]4->["\escale{$z_4$}"]5->["\escale{$k$}"]6;  
			};
\end{scope}
\begin{scope}[yshift=-3.6cm,nodes={set=import nodes}]
		\coordinate (bot4) at (0,0);
		\node (bot4) at (bot4) {\nscale{$\bot_4$}};
		\foreach \i in {0,...,7} { \coordinate (\i) at (\i*1.4cm+2cm,0) ;}
		\foreach \i in {3} {\fill[green!20, rounded corners] (\i) +(-0.4,-0.25) rectangle +(1.8,0.4);}
		\foreach \i in {2} {\fill[red!20, rounded corners] (\i) +(-0.4,-0.25) rectangle +(0.4,0.4);}
		\foreach \i in {0,...,7} { \node (\i) at (\i) {\nscale{$t_{4,\i}$}};}
\graph {
	(import nodes);
			bot4->[snake]0 ->["\escale{$k$}"]1->["\escale{$z_3$}"]2->["\escale{$X_{1}$}"]3->["\escale{$X_{3}$}"]4->["\escale{$X_{4}$}"]5->["\escale{$z_4$}"]6->["\escale{$k$}"]7;  
			};
\end{scope}
\begin{scope}[yshift=-4.8cm,nodes={set=import nodes}]
		\foreach \i in {0,...,4} { \coordinate (\i) at (\i*1.4cm,0) ;}
		\foreach \i in {0,...,4} { \coordinate (dot\i) at (\i*1.4cm,-0.3) ;}
		\foreach \i in {0,...,4} { \pgfmathparse{int(\i+5)} \node (bot\pgfmathresult) at (\i) {\nscale{$\bot_{\pgfmathresult}$}};}
		\foreach \i in {dot0,dot1,dot2,dot3,dot4} {  \node (\i) at (\i) {\nscale{$\vdots$}};}
\graph {
	(import nodes);
			};
\end{scope}
\path (bot1) edge [->] node[left] {\nscale{$\ominus_1$} } (bot2);
\path (bot2) edge [->] node[left] {\nscale{$\ominus_2$} } (bot3);
\path (bot3) edge [->] node[left] {\nscale{$\ominus_3$} } (bot4);
\path (bot4) edge [->] node[left] {\nscale{$\ominus_4$} } (bot5);
\path (bot5) edge [->] node[above] {\nscale{$\ominus_5$} } (bot6);
\path (bot6) edge [->] node[above] {\nscale{$\ominus_6$} } (bot7);
\path (bot7) edge [->] node[above] {\nscale{$\ominus_7$} } (bot8);
\path (bot8) edge [->] node[above] {\nscale{$\ominus_8$} } (bot9);
\end{tikzpicture}
\end{center}
\caption{A snippet of $A^\tau_I$ ($\tau=\{\nop,\inp,\res,\swap\}$) built from Example~\ref{ex:hitting_set} and showing the gadgets $T_1,\dots, T_4$.
Red colored area: the region $R=(sup, sig)$ where $sup(\bot_1)=0$; $sig(X_1)=\inp$; $sig(z_3)=\swap$; $sig(e)=\nop$ for all $e\in E(A^\tau_I)\setminus\{z_3,X_1\}$. 
Green colored area: the region $R=(sup, sig)$ where $sup(\bot_1)=0$; $sig(X_4)=\inp$; $sig(X_1)=\swap$; $sig(e)=\nop$ for all $e\in E(A^\tau_I)\setminus\{X_1, X_4\}$. 
}
\label{fig:nop_inp_res_swap_1}
\end{minipage}
\begin{minipage}{\textwidth}
\begin{center}
\begin{tikzpicture}[new set = import nodes]
\begin{scope}[nodes={set=import nodes}]%
		\coordinate (bot1) at (0,0);
		\node (bot1) at (bot1) {\nscale{$\bot_1$}};
		\foreach \i in {0,...,6} { \coordinate (\i) at (\i*1.4cm+2cm,0) ;}
		\foreach \i in {3} {\fill[green!20, rounded corners] (\i) +(-0.4,-0.25) rectangle +(4.6,0.4);}
		\foreach \i in {0,...,6} { \node (\i) at (\i) {\nscale{$t_{1,\i}$}};}
\graph {
	(import nodes);
			bot1->[snake]0 ->["\escale{$k$}"]1->["\escale{$z_3$}"]2->["\escale{$X_{1}$}"]3->["\escale{$X_{2}$}"]4->["\escale{$z_4$}"]5->["\escale{$k$}"]6;  
			};
\end{scope}
\begin{scope}[yshift=-1.2cm,nodes={set=import nodes}]%
		\coordinate (bot2) at (0,0);
		\coordinate (y1) at (1,0);
		\coordinate (y2) at (2,0);
		\node (bot2) at (bot2) {\nscale{$\bot_2$}};
		\node (y1) at (y1) {};
		\node (y2) at (y2) {};
		\foreach \i in {0,...,6} { \coordinate (\i) at (\i*1.4cm+3cm,0) ;}
		\foreach \i in {y2} {\fill[green!20, rounded corners] (\i) +(-0.4,-0.25) rectangle +(5.65,0.4);}
		\foreach \i in {0,...,6} { \node (\i) at (\i) {\nscale{$t_{2,\i}$}};}
\graph {
	(import nodes);
			bot2->[snake]y1 ->["\escale{$y_2$}"]y2->[snake]0->["\escale{$k$}"]1->["\escale{$z_3$}"]2->["\escale{$X_{2}$}"]3->["\escale{$X_{3}$}"]4->["\escale{$z_4$}"]5->["\escale{$k$}"]6;  
			};
\end{scope}
\begin{scope}[yshift=-2.4cm,nodes={set=import nodes}]%
		\coordinate (bot3) at (0,0);
		\node (bot3) at (bot3) {\nscale{$\bot_3$}};
		\foreach \i in {0,...,6} { \coordinate (\i) at (\i*1.4cm+2cm,0) ;}
		\foreach \i in {3} {\fill[green!20, rounded corners] (\i) +(-0.4,-0.25) rectangle +(4.5,0.4);}
		\foreach \i in {0,...,6} { \node (\i) at (\i) {\nscale{$t_{3,\i}$}};}
\graph {
	(import nodes);
			bot3->[snake]0 ->["\escale{$k$}"]1->["\escale{$z_3$}"]2->["\escale{$X_{1}$}"]3->["\escale{$X_{4}$}"]4->["\escale{$z_4$}"]5->["\escale{$k$}"]6;  
			};
\end{scope}
\begin{scope}[yshift=-3.6cm,nodes={set=import nodes}]%
		\coordinate (bot4) at (0,0);
		\node (bot4) at (bot4) {\nscale{$\bot_4$}};
		\foreach \i in {0,...,7} { \coordinate (\i) at (\i*1.4cm+2cm,0) ;}
		\foreach \i in {3} {\fill[green!20, rounded corners] (\i) +(-0.4,-0.25) rectangle +(0.4,0.4);}
		\foreach \i in {0,...,7} { \node (\i) at (\i) {\nscale{$t_{4,\i}$}};}
\graph {
	(import nodes);
			bot4->[snake]0 ->["\escale{$k$}"]1->["\escale{$z_3$}"]2->["\escale{$X_{1}$}"]3->["\escale{$X_{3}$}"]4->["\escale{$X_{4}$}"]5->["\escale{$z_4$}"]6->["\escale{$k$}"]7;  
			};
\end{scope}

\path (bot1) edge [->] node[left] {\nscale{$\ominus_1$} } (bot2);
\path (bot2) edge [->] node[left] {\nscale{$\ominus_2$} } (bot3);
\path (bot3) edge [->] node[left] {\nscale{$\ominus_3$} } (bot4);

\end{tikzpicture}
\end{center}
\caption{A sketch of the \enquote{first-glance} solution for $A^\tau_I$ ($\tau=\{\nop,\inp,\res,\swap\}$), where $I$ corresponds to Example~\ref{ex:hitting_set}.
Green colored area: the region $R=(sup, sig)$ where $sup(\bot_1)=0$; $sig(X_3)=\inp$; $sig(X_1)=sig(y_2)=\swap$; $sig(e)=\nop$ for all $e\in E(A^\tau_I)\setminus\{X_1, X_3,y_2\}$.} 
\label{fig:nop_inp_res_swap_2}
\end{minipage}
\begin{minipage}{\textwidth}
\begin{center}
\begin{tikzpicture}[new set = import nodes]
\begin{scope}[nodes={set=import nodes}]%
		\foreach \i in {0,...,2} { \coordinate (\i) at (\i*1.4cm+1.4,0) ;}
		\foreach \i in {1} {\fill[green!20, rounded corners] (\i) +(-0.4,-0.25) rectangle +(0.4,0.4);}
		\foreach \i in {0,...,2} { \node (\i) at (\i) {\nscale{$s^{i,j}_{i_1,\i}$}};}
\graph {
	(import nodes);
			0 ->["\escale{$v^{i,j}_1$}"]1->["\escale{$\oplus^{i,j}_1$}"]2;
			};
\end{scope}
\begin{scope}[yshift=-1.2cm,nodes={set=import nodes}]
		\foreach \i in {0,...,3} { \coordinate (\i) at (\i*1.4cm+1.4,0) ;}
		\foreach \i in {2} {\fill[green!20, rounded corners] (\i) +(-0.4,-0.25) rectangle +(0.4,0.4);}
		\foreach \i in {1} {\fill[blue!20, rounded corners] (\i) +(-0.4,-0.25) rectangle +(0.4,0.4);}
		\foreach \i in {0,...,3} { \node (\i) at (\i) {\nscale{$s^{i,j}_{i_2,\i}$}};}
\graph {
	(import nodes);
			0 ->["\escale{$v^{i,j}_2$}"]1->["\escale{$\oplus^{i,j}_2$}"]2->["\escale{$\oplus^{i,j}_1$}"]3;
			};
\end{scope}
\begin{scope}[yshift=-2.4cm,nodes={set=import nodes}]%
		\foreach \i in {0,...,4} { \coordinate (\i) at (\i*1.4cm+1.4,0) ;}
		\foreach \i in {3} {\fill[green!20, rounded corners] (\i) +(-0.4,-0.25) rectangle +(0.4,0.4);}
		\foreach \i in {2} {\fill[blue!20, rounded corners] (\i) +(-0.4,-0.25) rectangle +(0.4,0.4);}
		\foreach \i in {0,...,4} { \node (\i) at (\i) {\nscale{$s^{i,j}_{i_3,\i}$}};}
\graph {
	(import nodes);
			0 ->["\escale{$v^{i,j}_3$}"]1->["\escale{$\oplus^{i,j}_3$}"]2->["\escale{$\oplus^{i,j}_2$}"]3->["\escale{$\oplus^{i,j}_1$}"]4; 
			};
\end{scope}
\begin{scope}[yshift=-3.6cm,nodes={set=import nodes}]%
		
		\coordinate (init) at (0,0);
		\coordinate (init2) at (1.4cm,0);
		\coordinate (init3) at (2.8cm,0);
		\coordinate (init4) at (4.1cm,0);
		\coordinate (init5) at (4.6cm,0);
		\node(dots) at (0,0.6)  {\nscale{$\vdots$}};
		\node (init) at (init) {\nscale{$s^{i,j}_{i_\ell,0}$}};
		\node (init2) at (init2) {\nscale{$s^{i,j}_{i_\ell,1}$}};
		\node (init3) at (init3) {\nscale{$s^{i,j}_{i_\ell,2}$}};
		\node (init5) at (init5) {};
		\node (init5) at (init5) {\nscale{$\dots$}};
		\foreach \i in {0,...,4} { \coordinate (\i) at (\i*1.6cm+4.8cm,0) ;}
		\foreach \i in {3} {\fill[green!20, rounded corners] (\i) +(-0.4,-0.25) rectangle +(0.4,0.4);}
		\foreach \i in {2} {\fill[blue!20, rounded corners] (\i) +(-0.4,-0.25) rectangle +(0.4,0.4);}
		\node (0) at (0) {};
		\node (1) at (1) {\nscale{$s^{i,j}_{i_\ell,\ell-2}$}};
		\node (2) at (2) {\nscale{$s^{i,j}_{i_\ell,\ell-1}$}};
		\node (3) at (3) {\nscale{$s^{i,j}_{i_\ell,\ell}$}};
		\node (4) at (4) {\nscale{$s^{i,j}_{i_\ell,\ell+1}$}};
\graph {
	(import nodes);
			init ->["\escale{$v^{i,j}_\ell$}"]init2->["\escale{$\oplus^{i,j}_\ell$}"]init3->["\escale{$\oplus^{i,j}_{\ell-1}$}"]init4;
			0 ->["\escale{$\oplus^{i,j}_4$}"]1->["\escale{$\oplus^{i,j}_3$}"]2->["\escale{$\oplus^{i,j}_2$}"]3->["\escale{$\oplus^{i,j}_1$}"]4;
			};
\end{scope}
\end{tikzpicture}
\end{center}
\caption{The pyramidal approach of the relevant paths ensures that $\oplus$-events are solvable by regions independent of the size of $(\mathfrak{U}, M,\kappa)$.
Green colored area: a region $R=(sup, sig)$ solving $(\oplus^{i,j}_1, s)$ for all relevant $s\in S(A^\tau_I)$:
$sup(\bot_1)=0$;
for all $e\in E(A^\tau_I)$, if $e=\oplus^{i,j}_1$, then $sig(e)=\inp$;
if $e\in \{v^{i,j}_1, \oplus^{i,j}_2\}$, then $sig(e)=\swap$; otherwise $sig(e)=\nop$.
Blue colored area: a corresponding region solving $\oplus^{i,j}_2$.
These regions are independent from the positions of $G_{i_1},\dots, G_{i_\ell}$ in $A^\tau_I$ or $P_{i_n}$ in $G_{i_n}$, where $n\in \{1,\dots, \ell\}$.}
\label{fig:pyramidal_approach}
\end{minipage}
\end{figure}
\textbf{The Joining of $A^\tau_I$ by Relevant Paths.} 
Similar to the previous reductions, we essentially want to connect all gadgets by a simple directed path on which every event occurs exactly once.
However, since we want to ensure that if $\alpha$ is $\tau$-solvable then all (E)SSP atoms of $A^\tau_I$ are also $\tau$-solvable (by $d$-restricted regions), this is not directly possible for the gadgets $T_1,\dots, T_m$.
Instead, we complete the construction of $A^\tau_I$ through two further steps.
Firstly, for all $i\in \{1,\dots, m\}$, we extend the gadget $T_i$ to a (path-) gadget $G_i=\bot_i\sedge{}T_i$ with starting state $\bot_i$. 
Secondly, we use the events $\ominus_1,\dots, \ominus_{m+4}$ and connect the gadgets $G_1,\dots, G_m$ and $H_{0},\dots, H_{4}$ by $\bot_{1}\edge{\ominus_1}\bot_{2}\edge{\ominus_{2}}\dots\edge{\ominus_{m+4}}\bot_{m+5}$. 
The resulting TS is $A^\tau_I$, and its initial state is $\bot_1$.
Before we introduce the definition of $G_i$, in the following, we briefly outline which obstacles arise and, in order to overcome them, in which way they lead to $G_i$.

Let $i\in \{1,\dots, m\}$ and $\ell\in \{1,\dots, m_i\}$ be arbitrary but fixed.
Similar to the approach of region $R^{X,2}_{i,\ell}$ of Theorem~\ref{the:w2_result}.\ref{the:w2_result_nop_inp_set}, which is sketched for $i=3$ and $\ell=2$ by Figure~\ref{fig:nop_inp_set}, our aim is to solve $X_{i_\ell}$ \enquote{gadget-wise}.
In particular, to solve $(X_{i_\ell}, s)$ for all predecessor states $s$ of $t_{i,\ell+1}$ in $G_i$, that is, $\bot_i,\dots, t_{i,\ell}$, we want to construct a region $R=(sup, sig)$ such that as few events as possible are not mapped to $\nop$.
(Independent of $A^\tau_I$'s size, the region $R^{2,X}_{i,\ell}$ of Theorem~\ref{the:w2_result}.\ref{the:w2_result_nop_inp_set} maps four events not to \nop.)
First of all, look at the following definition:
$sup(\bot_1)=0$; 
for all $e\in E(A^\tau_I)$, if $e=X_{i_\ell}$, then $sig(e)=\inp$;
if $e$ is $X_{i_\ell}$'s direct predecessor, that is, $\edge{e}t_{i,\ell+1}$, then $sig(e)=\swap$;
otherwise $sig(e)=\nop$. 
In Figure~\ref{fig:nop_inp_res_swap_1}, the red colored area sketches this region for $X_{1_1}=X_1$ and its direct predecessor $z_3$;
the green colored area sketches this region for $X_{3_2}=X_4$ and its direct predecessor $X_1$.
Actually, $R$ is always well defined if $X_{i_\ell}\in E(T_j)$ implies that $X_{i_\ell}$'s direct predecessor $\edge{e}t_{i,\ell+1}$ also belongs to $E(T_j)$.
This is not true if there is an occurrence of $X_{i_\ell}$ in a gadget $T_j$, say at $t_{j,\ell'}$, such that $X_{i_\ell}$'s predecessor does not belong to $T_j$'s event set. 
For example, consider in Figure~\ref{fig:nop_inp_res_swap_1} the event $X_{4_2}=X_3$ of $T_4$ that occurs as $X_{2_2}$ in $T_2$.
In $T_4$, $X_3$ is directly preceded by $X_1$, but $X_1$ does not occur in $T_2$.
The following problem arises.
Since $sig(X_{i_\ell})=\inp$, there has to be an event $e$ on the unambiguous path $\bot_1\edge{}\dots\edge{}t_{j,\ell'}$ such that $sig(e)=\swap$.
Otherwise, $X_{i_\ell}$'s source $t_{i,\ell'}$ in $T_j$ would not satisfy $sup(t_{i,\ell'})=1$.
At first glance, a possible solution might be to implement an additional (unique) event $y_j$ on the path $\bot_j\sedge{}t_{j,0}$ for all $j\in \{1,\dots, m\}$ where $X_{i_\ell}$ belongs to $E(T_j)$ but $X_{i_\ell}$'s direct predecessor event does not. 
Then we would modify the region $R=(sup, sig)$ in a way, that $sig(y_j)=\swap$ for all relevant $j$.
Figure~\ref{fig:nop_inp_res_swap_2} sketches the situation for $y_2$.

Unfortunately, for this construction and the sketched region, $\vert \{e\in E(A^\tau_I)\mid sig(e)\not=\nop\}\vert \geq n+2$ holds, where $n$ is the number of gadgets in which $X_{i_\ell}$ occurs but its predecessor does not. 
Since $X_{i_\ell}$ could occur in numerous sets, in general, $n$ depends on the size of $M$ and does not necessarily respect the parameter $d$.
Thus, this approach yields not a parameterized reduction.
The next inelaborate solution to overcome this obstacle is to ensure that there is the same event, say $y$, on every path $\bot_j\sedge{}t_{j,0}$ for all $j\in \{1,\dots, m\}\setminus\{i\}$ such that $X_{i_\ell}\in E(T_j)$ but $X_{i_\ell}$'s predecessor is not in $E(T_j)$. 
However, one has to ensure that the already discussed difficulties are not transferred from $X_{i_\ell}$ to $y$.
Our solution uses \emph{relevant paths} to realize a pyramidal approach that is sketched by Figure~\ref{fig:pyramidal_approach}.
Instead of one single event $y$ (whose role is played by $\oplus^{i,j}_1$ in Figure~\ref{fig:pyramidal_approach}), this approach implements for every corresponding $T_j$ a unique directed path. 

Let $i\in \{1,\dots, m\}$ be arbitrary but fixed.
We extend the gadget $T_i$ to $G_i=\bot_i\edge{w_i}P_i\edge{u_i}T_i$ with starting state $\bot_i$ and events $w_i,u_i$ that embrace the path $P_i$, to be defined next. 
To be able to refer uniformly to the events $X_{i_1},\dots, X_{i_{m_i}}$ and $z_4$, we define $e^i_1=X_{i_1},\dots, e^i_{m_i}=X_{i_{m_i}}$ and $e^i_{m_i+1}=z_4$.
Let $j\in \{2,\dots, m_i+1\}$ be arbitrary but fixed and let $i_1<\dots < i_\ell \in \{1,\dots, m\}\setminus\{i\}$ be exactly the indices different from $i$ such that for the gadgets $T_{i_1},\dots, T_{i_\ell}$ 
we have $e^i_j\in E(T_{i_n})$ and $e^i_{j-1}\not\in E(T_{i_n})$, for all $n\in \{1,\dots,\ell\}$. 
For all $n\in \{1,\dots, \ell\}$, we say that $e^i_j$ is relevant for $G_{i_n}$ and 
\[
P^{i,j}_{i_n,n}= s^{i,j}_{i_n,0}\lEdge{v^{i,j}_n}s^{i,j}_{i_n,1}\lEdge{\oplus^{i,j}_n}s^{i,j}_{i_n,2}\lEdge{\oplus^{i,j}_{n-1}}\dots \lEdge{\oplus^{i,j}_1}s^{i,j}_{i_n,n+1}
\]
is the relevant path of $G_{i_n}$ that originates from $e^i_j$.
\begin{example}\label{ex:relevant_path}%
The event $e^1_3=z_4$ of $T_1$ of Figure~\ref{fig:nop_inp_res_swap_1} is preceded by $e^1_2=X_2$.
While the event $z_4$ occurs in $T_2,T_3$ and $T_4$, the event $X_2$ occurs in $T_2$ but not in $T_3$ and not in $T_4$.
Thus, $e^1_3$ is (only) relevant for $T_3=T_{i_1}$ and $T_4=T_{i_2}$, where $i_1=3$ and $i_2=4$.
The corresponding relevant paths are 
\[
P^{1,3}_{3,1}=s^{1,3}_{3,0}\lEdge{v^{1,3}_1}s^{1,3}_{3,1}\lEdge{\oplus^{1,3}_1}s^{1,3}_{3,2} \text{ and }
P^{1,3}_{4,2}=s^{1,3}_{4,0}\lEdge{v^{1,3}_2}s^{1,3}_{4,1}\lEdge{\oplus^{1,3}_2}s^{1,3}_{4,2}\lEdge{\oplus^{1,3}_1}s^{1,3}_{4,3} . 
\]
\end{example}%
Equipped with these definitions, we are prepared to define the gadget $G_i$.
If there are no relevant events for $G_i$, then $G_i=\bot_i\edge{w_i}q_i\edge{u_i}T_i$.
In particular, $P_i=q_i$.
Otherwise, let $e^{i_1}_{j_1},\dots, e^{i_n}_{j_n}$ be the events that are relevant for $G_i$ where $i_1\leq i_2\leq\dots\leq i_n$ and $j_1\leq j_2\leq\dots\leq j_n$.
Let $P^{i_1,j_1}_{i,\ell_1}, P^{i_2,j_2}_{i,\ell_2},\dots, P^{i_n,j_n}_{i,\ell_n}$ be the relevant paths of $G_i$ that origin from $e^{i_1}_{j_1},\dots, e^{i_n}_{j_n}$, respectively.
The path $P_i$ then originates from $G_i$'s relevant paths:
\[
G_i=\bot_i\lEdge{w_i}P^{i_1,j_1}_{i,\ell_1}\lEdge{c^i_1}P^{i_2,j_2}_{i,\ell_2}\lEdge{c^i_2}\dots\lEdge{c^i_n}P^{i_n,j_n}_{i,\ell_n}\lEdge{u_i}T_i
\]
See \cite{DBLP:journals/corr/abs-2007-12372} for a full  example.

\textbf{Theorem~\ref{the:w2_result}.\ref{the:w2_result_nop_inp_res_swap}: The $\tau$-Solvability of $\alpha$ Implies a Hitting Set.}
Let $R=(sup, sig)$ be a $d$-restricted $\tau$-region of $A^\tau_I$ that solves $\alpha$.
Since $R$ solves $\alpha$, one easily finds that $sig(k)=\inp$ and $sup(h_{0,2})=0$.
By $sig(k)=\inp$, we have $sup(h_{0,3})=1$; 
and $sup(h_{0,2})=0$ implies $sig(o_2)=\swap$. 
Moreover, by $sig(k)=\inp$ and $sig(o_2)=\swap$, we obtain that $sup(h_{1,1})=sup(h_{1,2})=sup(h_{2,1})=sup(h_{2,2})=0$.
This implies $sig(z_1),sig(z_2)\in \{\nop,\res\}$.
By $sig(k)=\inp$ and $sig(z_1),sig(z_2)\in \{\nop,\res\}$, we get $sup(h_{3,2})=sup(h_{4,2})=0$ and $sup(h_{3,3})=sup(h_{4,3})=1$.
This implies $sig(z_3)=sig(z_4)=\swap$.
Since $d=\kappa+4$ and $R$ is $d$-restricted, there are at most $\kappa$ events left whose signature is different from $\nop$.
Let $i\in \{1,\dots, m\}$ be arbitrary but fixed.
By $sig(k)=\inp$, we get $sup(t_{i,1})=0$ and $sup(t_{i,m_i+3})=1$.
Moreover, by $sig(z_3)=sig(z_4)=\swap$, we get $sup(t_{i,2})=1$ and $sup(t_{i,m_i+2})=0$.
Thus, there is an event $X\in E(T_i)$ such that $sig(X)\in \{\inp,\res,\swap\}$.
Since $i$ was arbitrary and $R$ is $d$-restricted, the set $S=\{X\in \mathfrak{U}\mid sig(X)\not=\nop\}$ is a sought-for HS of $I$.

\textbf{Theorem~\ref{the:w2_result}.\ref{the:w2_result_nop_inp_res_swap}: A Hitting Set Implies the $\tau$-Solvability of $A^\tau_I$.}
We argue for the $\tau$-solvability of $k$, implying the $\tau$-solvability of $\alpha$.
The following $d$-restricted $\tau$-region $R=(sup, sig)$ solves $\alpha$ and solves $(k,s)$ for all relevant $s\in \bigcup_{i=1}^mS(H_i)\setminus\{\bot_{m+1},\dots, \bot_{m+5}\}$, too:
$sup(\bot_1)=1$;
for all $e\in E(A^\tau_I)$, if $e=k$, then $sig(e)=\inp$;
if $e\in \{o_2,z_3,z_4\}$, then $sig(e)=\swap$;
if $e\in S$, then $sig(e)=\res$;
otherwise, $sig(e)=\nop$.

Let $i\in \{1,\dots, m\}$ be arbitrary but fixed.
The following region $R=(sup, sig)$ solves $(k,s)$ for all relevant $s\in S(G_i)$:
If $i=1$, then $sup(\bot_1)=0$, otherwise $sup(\bot_1)=1$;
for all $e\in E(A^\tau_I)$, if $e\in \{k,\ominus_{i-1}\}$, then $sig(k)=\inp$;
if $e\in \{\ominus_i, o_1,z_1, z_2, z_4\}$, then $sig(e)=\swap$;
if $e=z_3$, then $sig(e)=\res$;
otherwise, $sig(e)=\nop$.
It is easy to see that, for any $s\in\{\bot_{m+1},\dots, \bot_{m+5}\}$, this region can be modified to a $d$-restricted region that  solves $(k,s)$.

Let $i\in \{1,\dots, m_i\}$ be arbitrary but fixed.
The separability of $X_{i_1},\dots, X_{i_{m_i}}, z_4$ in $G_i$ has already been sketched in the explanation of the relevant paths.
Clearly, these events are separable in the gadgets in which they do not occur.
Also the helper events of the relevant paths are separable.
We omit the proofs for the sake of readability.

\section{Conclusion}\label{sct:concl}%
In this paper, we investigate the parameterized complexity of DR$\tau$S parameterized by $d$ and show $W[2]$-hardness for a range of Boolean types.
As a result, $d$ is ruled out for fpt-approaches for the considered types of nets.
As future work, it remains to classify DR$\tau$S exactly in the $W$-hierarchy.
Moreover, one may look for other more promising parameters:
If $N=(P, T, M_0, f)$ is a Boolean net, $p\in P$ and if the \emph{occupation number} $o_p$ of $p$ is defined by $o_p=\vert \{ M\in RS(N) \mid M(p)=1\} \vert $ then the \emph{occupation number} $o_N$ of $N$ is defined by $o_N=\text{max}\{ o_p \mid p\in P\}$.
If $\mathcal{R}$ is a $\tau$-admissible set (of a TS $A$) and $R\in \mathcal{R}$, then the support of $R$ determines the number of markings of $N_A^{\mathcal{R}}$ that occupy $R$, that, is, $o_R=\vert \{s \in S(A) \mid sup(s)=1 \}\vert$.
Thus, searching for a $\tau$-net where $o_N\leq n$, $n\in \mathbb{N}$, corresponds to searching for a $\tau$-admissible set $\mathcal{R}$ such that $\vert \{s \in S(A) \mid sup(s)=1\}  \vert \leq n$ for all $R\in \mathcal{R}$.
As a result, for each (E)SSP atom $\alpha$ there are at most $ \mathcal{O}(\binom{\vert S\vert}{o_N})$ fitting supports for $\tau$-regions solving $\alpha$.
Thus, the corresponding problem \emph{$o_N$-restricted $\tau$-synthesis} parameterized by $o_N$ is in XP if, in a certain sense, $\tau$-regions are fully determined by a given support $sup$.

\bibliography{myBibliography}%

\end{document}